\documentclass[journal]{IEEEtran}

\usepackage{ucs}
\usepackage[utf8x]{inputenc}
\usepackage[cmex10]{amsmath}
\usepackage{cite, amsfonts, amssymb, amsthm, bm, bbm, graphicx, relsize, multirow, booktabs, color, blindtext}
\usepackage[american]{babel}
\usepackage[T1]{fontenc}

\theoremstyle{definition}

\usepackage{algorithmic, algorithm}
\algsetup{indent=2em, linenodelimiter=.}

\theoremstyle{remark}

\DeclareMathOperator{\card}{card}
\DeclareMathOperator{\diag}{diag}
\DeclareMathOperator{\trace}{Tr}
\DeclareMathOperator*{\argmax}{\arg\,\max}

\IEEEoverridecommandlockouts

\title{Joint Design of surveillance radar and MIMO communication in cluttered environments}

\author{Emanuele~Grossi,~\IEEEmembership{Senior Member,~IEEE,}
 Marco~Lops,~\IEEEmembership{Fellow,~IEEE,}
 and Luca~Venturino,~\IEEEmembership{Senior Member,~IEEE}%
\thanks{E. Grossi and L. Venturino are with the Department
of Electrical and Information Engineering (DIEI), Universit\`a degli Studi di Cassino e del Lazio Meridionale, Italy 03043; e-mail: e.grossi@unicas.it, l.venturino@unicas.it.}%
\thanks{M. Lops is with the Department of Electrical Engineering and Information Technology (DIETI), Universit\`a degli Studi di Napoli ``Federico II,'' Italy 80125; email: lops@unina.it.}%
}

\begin{document}
\bstctlcite{BSTcontrol}
\maketitle

\begin{abstract}
In this study, we consider a spectrum sharing architecture, wherein a multiple-input multiple-output communication system cooperatively coexists with a surveillance radar. The degrees of freedom for system design are the transmit powers of both systems, the receive linear filters used for pulse compression and interference mitigation at the radar receiver, and the space-time communication codebook. The design criterion is the maximization of the mutual information between the input and output symbols of the communication system, subject to constraints aimed at safeguarding the radar performance. Unlike previous studies, we do not require any time-synchronization between the two systems, and we guarantee the radar performance on all of the range-azimuth cells of the patrolled region under signal-dependent (endogenous) and signal-independent (exogenous) interference. This leads to a non-convex problem, and an approximate solution is thus introduced using a block coordinate ascent method. A thorough analysis is provided to show the merits of the proposed approach and emphasize the inherent tradeoff among the achievable mutual information, the density of scatterers in the environment, and the number of protected radar cells. 
\end{abstract}

\begin{IEEEkeywords}
 Spectral coexistence, shared spectrum access for radar and communications (SSPARC); radar-communications convergence, joint system design, MIMO communications, surveillance radars, mutual information.
\end{IEEEkeywords}

\section{Introduction}

Spectral co-existence of sensing and communication systems, anticipated as a possible enabling technology for post-Fourth Generation (4G) wireless services in~\cite{Griffiths_2015}, has assumed more and more relevance with the deployment of Fifth Generation (5G)~\cite{Buzzi_2014}: for example, the December 2017 Third Generation Partnership Project (3GPP) first release of the 5G New Radio standard standardizes the usage of a 3~GHz carrier frequency to undertake single-carrier frequency-division multiple access in the uplink of terrestrial networks (see also~\cite{Gerzaguet_2017}). A foreseeable trend for Sixth Generation (6G) is an increased cell densification, with the transition from \emph{small cells} to \emph{tiny cells} and a corresponding transition from the Sub-6~GHz policy to the full utilization of C (4-8~GHz) and X (8-12~GHz) bandwidths~\cite{Saad_arxiv}. Since S, C, and X bands are traditionally assigned to sensing functions (and, very frequently, under military control in order to undertake surveillance tasks), co-existence between radar and wireless communications has become a necessity, more than an option. A significant recent development is the announcement of the shared spectrum access for radar and communications (SSPARC) program, by the defense advanced research projects agency (DARPA)~\cite{SSPARC_2013, Fitz_2014, Guerci_2015, Jacyna_2016}.

A number of approaches aimed at assessing the feasibility of full spectrum sharing have been proposed so far, the big divide being that among cooperative and un-cooperative architectures. Early studies (see, among the others,~\cite{Wang_2008, Deng_2013, Aubry_2014, Geng_2015, Hessar_2016, Raymond_2016, Singh_2018, Biswas_2018}) focused on the performance of a \emph{primary} radar system co-existing with un-licensed wireless users, while more recent results consider the performance of the communication system as the primary element of concern \cite{Aubry_2015, Bica_2016, Nartasilpa_2018, Zheng_2018_b}, and the radar waveform is carefully designed~\cite{Huang_2015, Meager_2016, Mahal_2017, Cheng_2018, Shi_2018}. Increased degrees of cooperation obviously allow safeguarding the performance of both the radar and the communication system. Such cooperative strategies avoid generating mutual interference through strict transmit policies coordination: for example, in~\cite{Blunt_2010, Chiryath_2016, Hassanien_2016, Grossi_2017, Grossi_2017_VTC, Grossi_2018_SP, Zhou_2019, Wang_Hassanien_2019, Zhang_2019} only one transmitter is active and carefully designed to ensure dual-function radar-communication, while in ~\cite{Mishra_2015, Stinco_2016, Cohen_2018} channel sensing techniques are borrowed from the large cognitive radio literature to detect and exploit spectral holes.

A different philosophy is the one proposed by~\cite{Li_Kumar_2016, Li_2016, Zheng_2017, Zheng_2018_a, Rihan_2018, Wang_2019}, wherein both systems are equipped with an active transmitter, but joint design (or co-design) of the radar waveform(s) and of the communication system codebook is undertaken. Generalizations of this approach to account for multiple-input multiple-output (MIMO) architectures of the radar and/or the communication systems, the presence of possible reverberation (clutter) induced by the radar transmitter onto the communication receiver, and the use of different radar waveform families have been recently considered~\cite{Li_2017, Qian_2018}. The major drawback of these approaches is that they assume complete freedom in the choice of the radar waveform: in fact, such waveforms cannot be chosen at will, but must comply with a number of requirements concerning resolution, variations in the signal modulus (amplifiers and A/D converters requires constant modulus signals), sidelobe level, and ambiguity~\cite{Skolnik_2008, Deng_2004, Li_2008}, whereby the performance achievable in practice---i.e., once the above requirements are translated into as many constraints in the joint design procedure---may be very distant from the theoretical ones, especially for what concerns the communication system~\cite{Zheng_2018_a, Chiryath_2019}. Moreover, the radar performance is guaranteed only at a specific resolution cell, thus making the design inappropriate in surveillance radars, where the monitored area is wide, and all the observed resolution cells should be protected by excessive interference. Finally, a high coordination between the two systems is needed, since they are required to operate in a time-synchronous manner~\cite{Zheng_2018_a}, and this may be too demanding or not allowed in some applications (e.g., for military radars).

In this context, starting from the preliminary results in~\cite{Grossi_2018}, we consider the problem of jointly designing the transceiver architecture of a surveillance radar and of a MIMO communication system, operating in full bandwidth overlap. The degrees of freedom for system optimization are, for the radar system, the transmit power and the receive filters, while, for the MIMO communication system, the whole codebook, whose space-time codewords (STC's) are bounded to span exactly one PRT. From the point of view of the radar system, whose transmit waveform is designed once and for all, this architecture replaces the concept of full cooperation---hardly feasible also due to security reasons---with that of \emph{awareness}. The design strategy is the maximization of the mutual information between the input and output symbols of the communication system, while the performance of the radar system is guaranteed by forcing the received signal-to-disturbance ratio (SDR) at each resolution bin to exceed a prescribed level. Major novelties of the present contribution are the detailed signal model, the global approach we take, in that \emph{all} the monitored radar resolution cells are protected, and the reduced degree of cooperation that is needed, since no time synchronization between the two systems is required. The resulting problem is very complex, whereby we propose a block coordinate ascent method (also known as alternating maximization) to find an approximate solution thereof. The analysis demonstrates that large gains with respect to a disjoint design are possible, especially in terms of achievable SDR's at the radar side, and that there is a fundamental tradeoff among the mutual information, the density of clutter scatterers in the surrounding environment, and the number of protected radar cells.

The rest of the paper is organized as follows. In the next section, the signal model at the radar and communication system sides is introduced. In Sec.~\ref{joint_opt_sec} the joint optimization problem is presented and (sub-optimally) solved, while in Sec.~\ref{num_ex_sec} a numerical example is provided to show the merits of the proposed strategy. Finally, some concluding remarks are given in Sec.~\ref{conclusion}.

\paragraph*{Notation} In the following, $\mathbb R$, $\mathbb R_+$, $\mathbb C$, and $\mathbb Z$ denotes the set of real, non-negative real, complex, and integer numbers, respectively; $x^+=\max\{x,0\}$ is the positive part of $x\in\mathbb R$, while $\bm x^+$ is the vector of positive parts of the entries of $\bm x\in\mathbb R^N$; $\bm x^*$, $\bm x^T$, and $\bm x^H$ denote the conjugate, the transpose, and the conjugate transpose of the vector $\bm x$; $\bm I_N$ is the $N\times N$ identity matrix; $\bm O_{N,M}$ is the $N\times M$ matrix with all zero entries; $\diag(a_1,\ldots,a_N)$ is the $N\times N$ diagonal matrix with entries $\{a_i\}_{i=1}^N$ on the principal diagonal; $[\bm X]_{a:b,c:d}$ is the sub-matrix consisting of the rows $a$ through $b$ and the columns $c$ through $b$ of the matrix $\bm X$; $\bm X^{-1/2}$ is the square root of the inverse of the Hermitian positive definite matrix $\bm X$; $\mathbb E [\, \cdot\,]$ denotes statistical expectation; $\mathcal N_c(\bm 0, \bm X)$ denotes the complex circularly-symmetric Gaussian distribution with covariance matrix $\bm X$; and $\mathbbm 1_\mathcal{A}$ is the indicator function of the condition $\mathcal A$, i.e., $\mathbbm 1_\mathcal{A}=1$, if $\mathcal A$ holds true, and $\mathbbm 1_\mathcal{A}=0$, otherwise.

\section{Signal model}

\begin{figure}[t]
\centering
\ifCLASSOPTIONtwocolumn
\centerline{\includegraphics[width=\columnwidth]{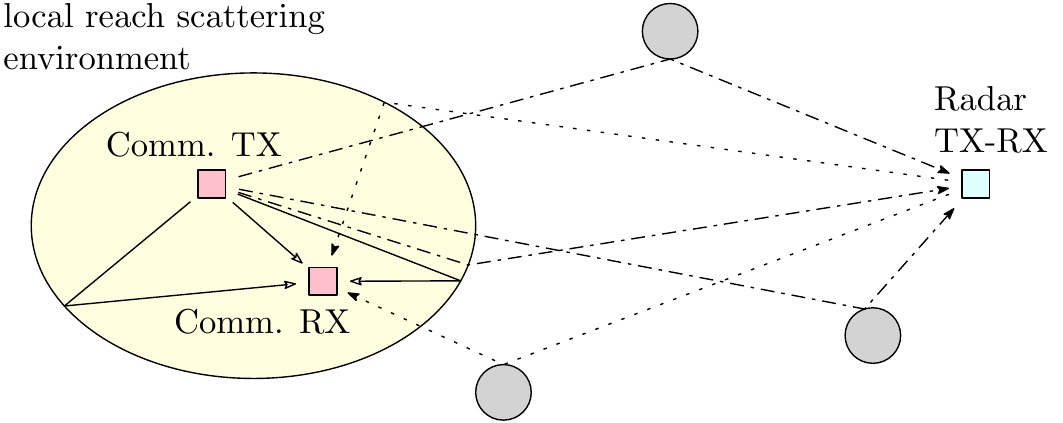}}
\else
\centerline{\includegraphics[width=0.7\textwidth]{fig_01.pdf}}
\fi
\caption{Considered scenario of coexistence between a MIMO communication system and a radar. \label{Fig-scheme}}
\end{figure}

We consider the scenario outlined in Fig.~\ref{Fig-scheme}, where a MIMO communication system coexists with a surveillance radar on the same bandwidth $W$. The communication system operates in a local rich scattering environment (e.g., an urban area), whose size is in the order of $c/W$, $c$ denoting the speed of light. The radar is located outside this area, and monitors a large region, that includes the one where the communication system operates; the range resolution of the radar is in the order of $c/(2W)$.

The radar is equipped with a non-scanning wide-beam transmit antenna and emits the following train of (encoded) pulses
\begin{equation}
 s(t)= \sqrt{P_r} \sum_{p\in\mathbb Z} \sum_{\ell=0}^{L-1}q(\ell)\phi(t-\ell T_c - pT)
\end{equation}
where:
\begin{itemize}
 \item $P_r$ is the average transmit power;
 \item $\bm q=(q(0)\, \cdots \,q(L-1))^T\in \mathbb C^L$ is the (fast-time) code sequence, used to modulate the subpulses composing each pulse, also called chip sequence; we set $\Vert \bm q \Vert^2=N$, with $N$ a positive integer greater than $L$;
 \item $T_c\approx 1/W$ is the chip period;
 \item $\phi(t)$ is the subpulse, a baseband waveform with support\footnote{We hasten to underline that the assumption on the support of $\phi(t)$ is made here just to simplify the exposition: the sufficient condition is that $\phi(t)$ be a Nyquist waveform, i.e., that $\int_{\mathbb R} \phi(t) \phi(t-\ell T_c)dt= 0$, for any $\ell\neq0$.} in $[0,T_c]$, bandwidth $W$, and such that $\int_{\mathbb R} |\phi(t)|^2dt=T_c$;
 \item $L$ is the number of subpulses in each pulse, so that $LT_c$ is the pulse duration; and
 \item $T=NT_c$ is the pulse repetition time (PRT).
 \end{itemize}
The number of \emph{non-ambiguous} range cells\footnote{Ambiguity arises from the periodicity of the radar waveform $s(t)$, and echoes whose arrival time differ for an integer multiple of $T$ (i.e., targets with ranges spaced by integer multiples of $cT/2$) are not distinguishable~\cite{Skolnik_2001}.} is $N\approx WT$, and is typically much larger than $L$.

The communication transmitter is equipped with $M$ omni-directional antennas, and the waveform emitted by antenna $m$ is
\begin{equation}
x_m(t)=\sum_{i\in\mathbb Z} c_m(i) \phi(t-iT_c)
\end{equation}
$m=1,\ldots,M$, where $c_m(i)$ is the symbol sent at epoch $i$. The symbols transmitted by the $M$ antennas during $N$ signaling intervals (i.e., $\{c_m(pN+i): m=1,\ldots, M, i=0,\ldots, N-1\}$, with $p$ an integer) form an $M\times N$ STC, which is drawn from a complex circularly-symmetric Gaussian (CCSG) codebook.

\begin{figure}[t]
 \centering
 \ifCLASSOPTIONtwocolumn
 \centerline{\includegraphics[width=0.7\columnwidth]{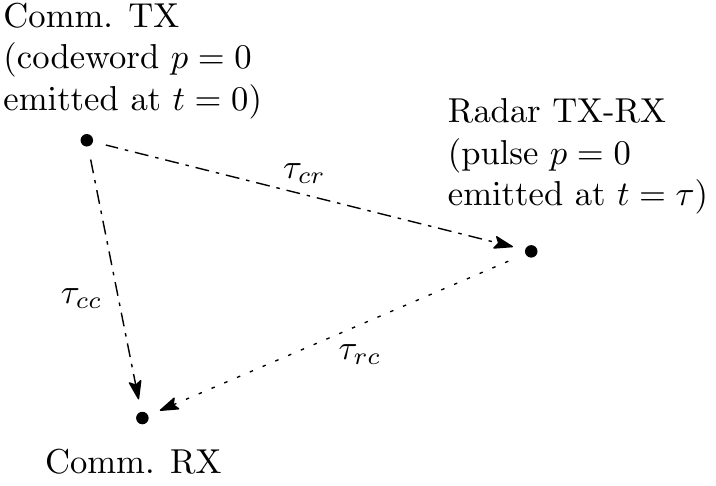}}
 \else
 \centerline{\includegraphics[width=0.45\textwidth]{fig_02.pdf}}
 \fi
\caption{Transmit and propagation delays between the two systems. \label{fig_02}}
\end{figure}

The time delay of the radar transmitter with respect to the communication transmitter is denoted by $\tau$; $\tau_{cc}$ is the propagation delay between the communication transmitter and receiver, while $\tau_{cr}$ and $\tau_{rc}$ are the (smallest) traveling time between communication transmitter and radar receiver and between radar transmitter and communication receiver, respectively (cfr. Fig.~\ref{fig_02}).

\subsection{Data receiver}

\begin{figure}[t]
 \centering
 \ifCLASSOPTIONtwocolumn
 \centerline{\includegraphics[width=1.04\columnwidth]{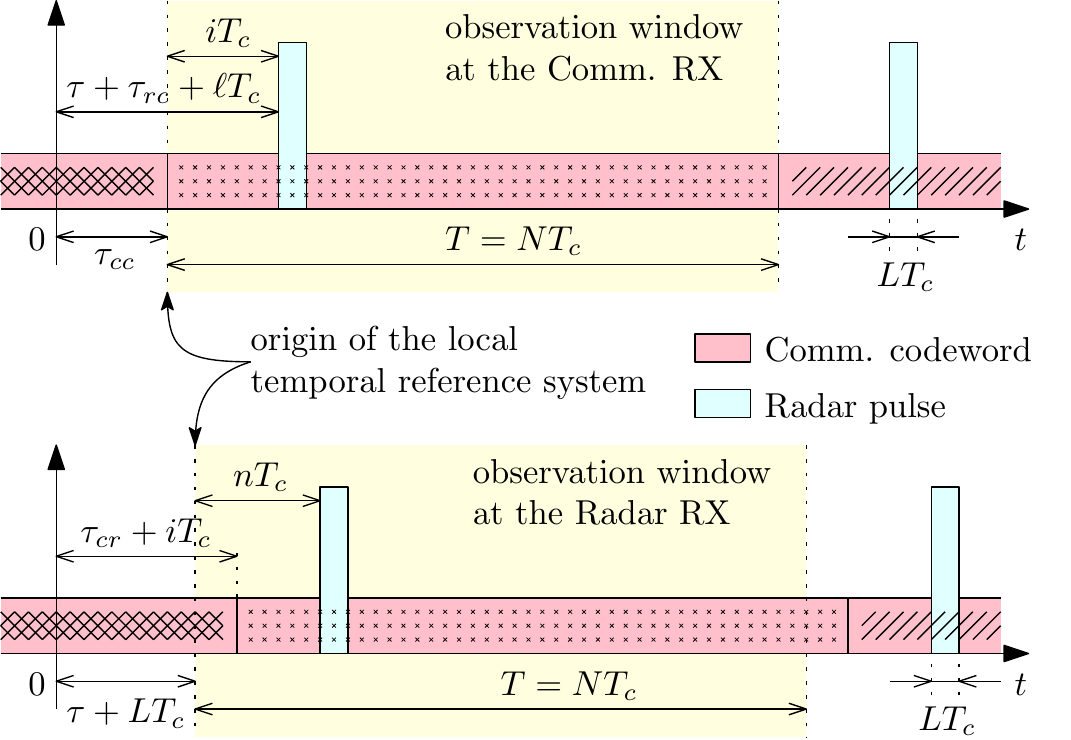}}
 \else
 \centerline{\includegraphics[width=0.7\textwidth]{fig_03.pdf}}
 \fi
\caption{Observation windows at the communication and radar receivers for codeword/PRT $p=0$: both the signal of interest and the interference from the coexisting system is present. \label{fig_03}}
\end{figure}

The communication receiver is equipped with $K$ omni-directional antennas and process the signal observed in the time interval corresponding to the $p$-th codeword, as shown in Fig.~\ref{fig_03}. The signal received by antenna $k$ (expressed with respect to the temporal reference system of the communication receiver) can be modeled as
\begin{equation}
r_k(t) =\underbrace{\sum_{m=1}^Mh_{k,m} x_m(t)}_{\text{signal of interest}} +\underbrace{\sum_{i=0}^{N-1} \alpha_{k,i} s(t-iT_c)}_{\text{radar inteference}} + \underbrace{v_k(t)}_{\substack{\text{thermal}\\ \text{noise}}} \label{comm_rx_signal}
\end{equation}
for $t\in\big[pT,(p+1)T\big]$, where:
\begin{itemize}
 \item $h_{k,m}\in \mathbb C$ is the gain of the channel linking the receive antenna $k$ to the transmit antenna $m$, assumed to be perfectly estimated and known at the transmitter;\footnote{We are assuming here that possible a Doppler shift between transmitter and receiver can be neglected over a time interval of length $T$, corresponding to the transmission of a STC.}
 \item $\alpha_{k,i}\in \mathbb C$ is the sum of the amplitudes of the radar echos hitting antenna $k$ at times\footnote{The radar interference is $\sum_{i=0}^{N-1} \sum_{d=0}^\infty \alpha_{k,i,d} s(t-iT_c-dT)$, where $\alpha_{k,i,d}$ is the amplitude of an echo hitting antenna $k$ at time $iT_c+dT$, and the term in~\eqref{comm_rx_signal} is obtained by defining $\alpha_{k,i}=\sum_{d=0}^\infty \alpha_{k,i,d}$ and by exploiting the periodicity of the radar waveform. \label{footnote_radar_interference}} $\{iT_c+dT\}_{d=0}^\infty$; we model $\bm \alpha_i= (\alpha_{1,i} \, \cdots \, \alpha_{K,i})^T\in\mathbb C^K$ as a CCSG random vector with covariance matrix $\bm \Sigma_{\alpha,i}$, perfectly estimated and known at the transmit side,\footnote{Remarkable examples for $\bm \Sigma_{\alpha,i}$ are the diagonal and the rank-one cases: in the former, there is a line-of sight component linking the scatterer (or the radar) and the receive antennas, while, in the latter, there is no line-of sight component, and independent rays from the rich scattering environment arrive at the receive antennas.} and assume that $\bm \alpha_i$ and $\bm \alpha_j$ are independent for $i\neq j$; and
 \item $v_k(t)$ is the thermal noise, modeled as a CCSG white process with power spectral density $\sigma^2_v$.
\end{itemize} 

Let $r_{k,i}= \frac{1}{T_c}\int_{\mathbb R} r_k(t) \phi^*(t-pT - iT_c)dt$, $i=0,\ldots,N-1$, be the projections of the signal received at the $k$-th antenna onto $N$ time-shifted versions of the baseband pulse (normalized by $T_c$) and $\bm c=(\bm c_1^T \, \cdots \, \bm c_M^T)^T \in \mathbb C^{MN}$ the $p$-th STC, where $\bm c_m=( c_m(0) \, \cdots \, c_m(N-1))^T \in \mathbb C^N$. Then, the discrete-time signal $\bm r_k=(r_{k,0}\, \cdots \, r_{k,N-1})^T\in \mathbb C^N$ can be written as 
\begin{equation}
\bm r_k= \big( (h_{k,1}\, \cdots \, h_{k,M}) \otimes \bm I_N\big) \bm c +\sqrt{P_r} \sum_{i=0}^{N-1} \alpha_{k,i} \bm q_i + \bm v_k. 
\end{equation}
where $\bm q_i\in\mathbb C^N$ is the vector obtained by taking a downwards circular shift of $i$ positions of $(\bm q^T \, 0 \, \cdots \, 0)^T \in \mathbb C^N$, and $\bm v_k \sim \mathcal N_c(\bm 0, \bm P_v \bm I_N)$ is the noise vector, with $P_v=\sigma^2_v/T_c$. Stacking the received signals in a unique vector $\bm r =(\bm r_1^T \, \cdots \, \bm r_K^T)^T\in \mathbb C^{KN}$, we obtain
\begin{equation}
\bm r = (\bm H \otimes \bm I_N) \bm c+ \sqrt{P_r} \sum_{i=0}^{N-1} \bm \alpha_i \otimes \bm q_i +\bm v
\end{equation}
where $\bm v=(\bm v_1^T\, \cdots \, \bm v_K^T)^T \in \mathbb C^{KM}$, and
\begin{equation}
 \bm H=\begin{pmatrix}
 h_{1,1} & \cdots & h_{1,M}\\
 \vdots & & \vdots\\
 h_{K,1} & \cdots & h_{K,M}
 \end{pmatrix} \in\mathbb C^{K \times M}.
\end{equation}

With this notation, the (normalized) mutual information between the received vector $\bm r$ and the STC $\bm c$ is
\ifCLASSOPTIONtwocolumn
\begin{align}
 R &= \frac1N\log \det \left(\vphantom{\left(P_r \sum_{i=0}^{N-1} \bm \Sigma_{\alpha,i} \otimes (\bm q_i\bm q_i^H) +P_v \bm I_{NK} \right)^{-1}} \bm I_{NK} + (\bm H \otimes \bm I_N) \bm C (\bm H \otimes \bm I_N)^H \right. \notag\\
 &\quad \left. \times \left(P_r \sum_{i=0}^{N-1} \bm \Sigma_{\alpha,i} \otimes (\bm q_i\bm q_i^H) +P_v \bm I_{NK} \right)^{-1}\right) \label{R_expr}
\end{align}
\else
 \begin{equation}
  R = \frac1N\log \det \left(\vphantom{\left(P_r \sum_{i=0}^{N-1} \bm \Sigma_{\alpha,i} \otimes (\bm q_i\bm q_i^H) +P_v \bm I_{NK} \right)^{-1}} \bm I_{NK} + (\bm H \otimes \bm I_N) \bm C (\bm H \otimes \bm I_N)^H \left(P_r \sum_{i=0}^{N-1} \bm \Sigma_{\alpha,i} \otimes (\bm q_i\bm q_i^H) +P_v \bm I_{NK} \right)^{-1}\right) \label{R_expr}
 \end{equation}
\fi
bits per channel use, where $\bm C=\mathbb E \left[ \bm c \bm c^H\right]$ is the covariance matrix of $\bm c$. Such mutual information represents an upper bound to the achievable transmission rate and can be approached provided $N$ is large enough.

\subsection{Radar receiver}

The radar forms $J$ simultaneous orthogonal beams (defining as many azimuth bins), so as to cover the area illuminated by the non-scanning wide transmit beam, and elaborates the signal received in the $p$-th PRT, as shown in Fig.~\ref{fig_03}. Assuming the presence of a point-like target with delay\footnote{Delays smaller than the duration of the emitted pulse correspond to distances of no interest for the radar.} $(L+n)T_c$, $n\in\{0,\dots,N-L\}$, in the $j$-th azimuth direction, $j\in\{1,\ldots, J\}$, the continuous-time signal received from the $j$-th azimuth beam (expressed with respect to the temporal reference system of the radar receiver) can be modeled as
\ifCLASSOPTIONtwocolumn
 \begin{align}
 y_j (t) &= \underbrace{g_{n,j} s(t-n T_c)}_{\text{target echo}}+ \underbrace{\sum_{i=0}^{N-1} \gamma_{i,j} s(t-iT_c)}_{\text{radar clutter}} \notag\\
 & \quad + \sum_{m=1}^M \sum_{i=0}^{N-1} \sum_{d=0}^\infty \beta_{m,i,j,d} \notag\\
 & \quad \quad \underbrace{\quad \quad \times x_m \big(t+\tau+LT_c- (\tau_{cr}+iT_c+dT) \big)}_{\text{data inteference}} \notag\\
 & \quad + \underbrace{u_j (t)}_{\substack{\text{thermal}\\ \text{noise}}} \label{radar_rx_signal}
 \end{align}
\else
 \begin{align}
 y (t) &= \underbrace{g_{n,j} s(t-n T_c)}_{\text{target echo}}+ \underbrace{\sum_{i=0}^{N-1} \gamma_i s(t-iT_c)}_{\text{clutter}} \notag\\
 & \quad + \underbrace{\sum_{m=1}^M \sum_{i=0}^{N-1} \sum_{d=0}^\infty \beta_{m,i,d} x_m \big(t+\tau+LT_c- (\tau_{cr}+iT_c+dT) \big)}_{\text{data inteference}} + \underbrace{u (t)}_{\substack{\text{thermal}\\ \text{noise}}} \label{radar_rx_signal}
 \end{align}
\fi
for $t\in \big[pT, (p+1)T\big]$, where:
\begin{itemize}
 \item $g_{n,j}\in \mathbb C$ is the amplitude of the target echo, modeled as a zero-mean random variable with variance $\sigma^2_{g,n,j}$;
 \item $\gamma_{i,j}\in \mathbb C$ is the sum of the amplitudes of the clutter echoes received at times\footnote{The comment in Footnote~\ref{footnote_radar_interference} is also valid here.} $\{iT_c+dT\}_{d=0}^\infty$ from the $j$-th azimuth bin, modeled as a zero-mean random variable with variance $\sigma^2_{\gamma,i,j}$; we assume that $\sigma^2_{\gamma,i,j}$ and $\sigma^2_{\gamma,i',j'}$ are independent for $(i,j)\neq (i',j')$, and that $\sigma^2_{\gamma,i,j}$ has been perfectly estimated;
 \item $\beta_{m,i,j,d}\in \mathbb C$ is the amplitude of the echo caused by the signal emitted by antenna $m$ of the communication system and hitting the radar from the $j$-th azimuth direction after a traveling time $\tau_{cr}+iT_c+dT-\tau-LT_c$; we model $\beta_{m,i,j,d}$ as a zero-mean random variable and assume that $\beta_{m,i,j,d}$ and $\beta_{m',i',j',d'}$ are independent for $(i,j,d)\neq(i',j',d')$; we also assume that\footnote{Two notable cases are those where $\sigma^2_{\beta,m,m',i,j}$ is independent of $(m,m')$, and where it is equal to zero for all $m\neq m'$. In the former case, there is a line-of-sight component linking the transmit antennas and the scatterer (or the radar), so that $\beta_{m,i,j,d}=\beta_{m',i,j,d}$; in the latter, there is no line-of-sight component, and independent rays from the rich scattering environment arrive at the radar.} $\sigma^2_{\beta,m,m',i,j}= \sum_{d=0}^\infty \mathbb E[ \beta_{m,i,j,d}\beta_{m',i,j,d}^*]$ has been perfectly estimated; and
 \item $u_j (t)$ is the additive thermal noise, modeled as a CCSG process with power spectral density $\sigma^2_u$. 
\end{itemize}

Letting $y_{i,j}= \frac{1}{T_c} \int_{\mathbb R} y_j(t) \phi^*(t-pT-iT_c) dt$, $i=0,\ldots,N-1$, be the projections of the signal received on the $j$-th azimuth bin during the $p$-th PRT, we obtain the $N$-dimensional discrete-time signal
\ifCLASSOPTIONtwocolumn
 \begin{align}
 \bm y_j&= \begin{pmatrix} y_{0,j} & \cdots & y_{N-1,j} \end{pmatrix}^T \notag\\
 &= \sqrt{P_r} g_n \bm q_n + \sqrt{P_r} \sum_{i=0}^{N-1} \gamma_{i,j} \bm q_i \notag\\
 & \quad + \sum_{m=1}^M \sum_{i=0}^{N-1} \sum_{d=0}^\infty \beta_{m,i,j,d} \bm c_{m,i,d} + \bm u_j
 \end{align}
\else
 \begin{align}
 \bm y_j&= \begin{pmatrix} y_{0,j} & \cdots & y_{N-1,j} \end{pmatrix}^T \notag\\
 &= \sqrt{P_r} g_n \bm q_n + \sqrt{P_r} \sum_{i=0}^{N-1} \gamma_{i,j} \bm q_i + \sum_{m=1}^M \sum_{i=0}^{N-1} \sum_{d=0}^\infty \beta_{m,i,j,d} \bm c_{m,i,j,d} + \bm u_j
 \end{align}
\fi
where $\bm u_j\in\mathcal N_c(\bm 0, \bm P_u \bm I_N)$, with $P_u=\sigma^2_u/T_c$, and
\begin{equation}
\bm c_{m,i,d}=\begin{pmatrix} c_m\big((p-d)N- \nu-i\big) \\ \vdots \\ c_m\big((p-d)N-\nu-i +N-1\big)\end{pmatrix}
\end{equation} 
is the sequence of $N$ symbols transmitted by antenna $m$ of the communication system that fall in the $p$-th PRT, with\footnote{Notice that the assumption that $\tau_{cr}-\tau$ be an integer is not necessary and is made here only to simplify description of the signal model.} $\nu=(\tau_{cr}-\tau-LT_c)/T_c\in\mathbb Z$ (cfr. Fig.~\ref{fig_03}). Notice that $\bm c_{m,i,d}$ contains the last $\ell_i=(\nu+i) \mod N$ symbols of the $m$-th segment of codeword $p-d- \big \lfloor (\nu+i)/N \big \rfloor-1$ and the first $N-\ell_i$ symbols of the $m$-th segment of codeword $p-d- \big \lfloor (\nu+i)/N \big \rfloor$. Hence, assuming that codewords emitted at different epochs are independent, the cross-covariance matrix of $\bm c_{m,i,d}$ and $\bm c_{m',i,d}$ is
\ifCLASSOPTIONtwocolumn
\begin{align}
&\bm C_{m,m',i}=\mathbb E\left[ \bm c_{m,i,d} \bm c_{m',i,d}^H\right] \notag\\
&= \begin{pmatrix}
[\bm C_{m,m'}]_{N-\ell_i+1:N,N-\ell_i+1:N} & & \bm O_{\ell_i,N-\ell_i} 
\\[10pt]
\bm O_{N-\ell_i,\ell_i} & & [\bm C_{m,m'}]_{1:N-\ell_i,1:N-\ell_i}
\end{pmatrix}
\end{align}
\else
 \begin{equation}
  \bm C_{m,m',i}=\mathbb E\left[ \bm c_{m,i,d} \bm c_{m',i,d}^H\right] = \begin{pmatrix}
[\bm C_{m,m'}]_{N-\ell_i+1:N,N-\ell_i+1:N} & & \bm O_{\ell_i,N-\ell_i} 
\\[10pt] \bm O_{N-\ell_i,\ell_i} & & [\bm C_{m,m'}]_{1:N-\ell_i,1:N-\ell_i}
\end{pmatrix}
 \end{equation}
\fi
where $\bm C_{m,m'}=E\left[ \bm c_m \bm c_m'^H\right]$. Interestingly, $\bm C_{m,m',i}$ can also be expressed as a linear function of the covariance matrix $\bm C$:
\begin{equation}
 \bm C_{m,m',i} = \bm A_{m,i} \bm C \bm A_{m',i}^T + \bm B_{m,i} \bm C \bm B_{m',i}^T
\end{equation}
where $\bm A_{m,i}, \bm B_{m,i}\in\mathbb C^{N\times MN}$ are defined as
\begin{subequations}
\begin{align}
 \bm A_{m,i}& = \begin{pmatrix}
 \bm O_{\ell_i,(m-1)N} & \bm O_{\ell_i,N-\ell_i} & \bm O_{\ell_i,(M-m)N+\ell_i} \\
 \bm O_{N-\ell_i, (m-1)N} & \bm I_{N-\ell_i} & \bm O_{N-\ell_i,(M-m)N+\ell_i}
 \end{pmatrix}\\
 \bm B_{m,i}& = \begin{pmatrix}
 \bm O_{\ell_i,mN-\ell_i} & \bm I_{\ell_i} & \bm O_{\ell_i,(M-m)N} \\
 \bm O_{N-\ell_i, mN-\ell_i} & \bm O_{N-\ell_i,\ell_i} & \bm O_{N-\ell_i,(M-m)N}
 \end{pmatrix}.
\end{align}%
\end{subequations}

Assume that pulse compression at range bin $n$ and azimuth bin $j$ is carried out by using the linear filter $\bm w_{n,j}$; then, the filter output is
\begin{align}
 \bm w_{n,j}^H\bm y_j &=\sqrt{P_r} g_n \bm w_{n,j}^H \bm q_n + \sqrt{P_r } \sum_{i=0}^{N-1} \gamma_{i,j} \bm w_{n,j}^H \bm q_i \notag\\
 &\quad + \sum_{m=1}^M \sum_{i=0}^{N-1} \sum_{d=0}^\infty \beta_{m,i,j,d} \bm w_{n,j}^H \bm c_{m,i,d} + \bm w_{n,j}^H \bm u
\end{align}
and the SDR at resolution cell $(n,j)\in \{0,\ldots, N-L\} \times \{1,\ldots,J\}$ is
\ifCLASSOPTIONtwocolumn
 \begin{align}
 \text{SDR}_{n,j}& =P_r \sigma^2_{g,n,j} |\bm w_{n,j}^H\bm q_n|^2 \left[\bm w_{n,j}^H \left( P_r\sum_{i=0}^{N-1} \sigma^2_{\gamma,i,j} \bm q_i \bm q_i^H \right. \right.\notag\\
 & \quad + \sum_{i=0}^{N-1} \sum_{m=1}^M \sum_{m'=1}^M \sigma^2_{\beta,m,m',i,j} (\bm A_{m,i} \bm C \bm A_{m',i}^T \notag\\
 & \quad \left. \left. \vphantom{\sum_{i=0}^{N-1}} + \bm B_{m,i} \bm C \bm B_{m',i}^T) +P_u\bm I_N \right)\bm w_{n,j}\right]^{-1}. \label{SDR_def}
 \end{align}
\else
 \begin{align}
 \text{SDR}_{n,j} &=P_r \sigma^2_{g,n,j} |\bm w_{n,j}^H\bm q_n|^2 \left[\bm w_{n,j}^H \left( P_r \sum_{i=0}^{N-1} \sigma^2_{\gamma,i,j} \bm q_i \bm q_i^H \right. \right.\notag\\
 & \quad \left. \left. + \sum_{i=0}^{N-1} \sum_{m=1}^M \sum_{m'=1}^M \sigma^2_{\beta,m,m',i,j} (\bm A_{m,i} \bm C \bm A_{m',i}^T + \bm B_{m,i} \bm C \bm B_{m',i}^T) +P_u\bm I_N \right)\bm w_{n,j}\right]^{-1}. \label{SDR_def}
 \end{align}
\fi

\subsection{Remarks on the signal model}

Since the radar is non-scanning and the time duration of the communication STC's is equal to the PRT of the radar, the second order statistics of the interference do not chance over consecutive PRT's/STC's, but remain constant on long time-scales, dictated by the time variability of the covariances of the scattering coefficients (the entries of the matrices $\bm \Sigma_{\alpha,i}$, at the communication system, the coefficients $\sigma^2_{\beta,m,m',i,j}$ at the radar side), as it can be seen from~\eqref{R_expr} and~\eqref{SDR_def}, respectively. Such slow time variability can be exploited in practice to obtain accurate estimates of these quantities, as it is already done for the channel matrix at the communication system and for the clutter covariance at the radar side. Remarkably, no time-synchronization between the two system is needed.

Concerning the interference model, Eqs.~\eqref{comm_rx_signal} and~\eqref{radar_rx_signal} are quite general and subsume the limiting cases of \emph{large scale} and \emph{small scale} reverberation phenomena. The former, is typically generated by physically extended obstacles that are in line-of-sight with the transmitter and the receiver. In this case, the interfering coefficients over the antennas---$\{\alpha_{k,i}\}_{k=1}^K$, concerning the interference from the radar to the communication receive array, and/or $\{\beta_{m,i,j,d}\}_{m=1}^M$, concerning the interference from the communication transmit array to the radar---are all equal. Small scale phenomena are instead generated when the communication system is \emph{embedded} in an urban area, and no line-of-sight between the communication system and any physically extended object is present: the clutter is generated \emph{locally} and the scatterers proximity to the transmit array (for the interference generated at the radar side) and/or to the receive array (for the interference generated at the communication system) explains the different aspect angles under which each receive antenna is seen. In this case case, the interfering coefficients over the antennas, $\{\alpha_{k,i}\}_{k=1}^K$ and/or $\{\beta_{m,i,j,d}\}_{m=1}^M$, are modeled as independent random variables.

\section{Joint optimization}\label{joint_opt_sec}

The radar is normally designed to guarantee a minimum level of SDR at each resolution bin, so as to be able to detect targets with specified radar cross-sections at specified locations; we denote $\rho_{n,j}$ such minimum required SDR at resolution cell $(n,j)\in \mathcal X$, where $\mathcal X\subseteq \{0,\ldots, N-L\} \times \{1,\ldots,J\}$ denotes the set of cells under observation. As to the communication system, high data rates are always desirable, and the covariance matrix of the STC's should be chosen to obtain a mutual information as large as possible, while not exceeding the available transmit power. We therefore propose to maximize the mutual information at the communication system with a constraint on the minimum required SDR at each radar cell under observation, and the optimization problem tackled here is
\begin{equation}
 \begin{aligned}
 \max_{\substack{\bm C \in \mathbb C^{MN\times MN} \\ \{ \bm w_{n,j}\}_{(n,j)\in\mathcal X} \in \mathbb C^N \\ P_r\in\mathbb R}} & \quad R ( \bm C, P_r )\\
 \text{s.t} & \quad \bm C \text{ Hermitian positive semi-definite}\\
 & \quad \frac1N \trace\{\bm C\} \leq P_{c,\text{max}}\\
 & \quad \text{SDR}_{n,j} (\bm C, P_r, \bm w_{n,j}) \geq \rho_{n,j}, \forall (n,j) \in \mathcal X\\
 & \quad 0\leq P_r \leq P_{r,\text{max}}
 \end{aligned} \label{opt_prob}
\end{equation}
where $P_{r,\text{max}}$ and $P_{c,\text{max}}$ is the maximum transmit power at the radar and communication system, respectively.

Clearly, this problem admits a solution if and only if the constraints can be satisfied at least when the communication system is not transmitting, the radar uses all the available power, and the radar receive filters are chosen so as to maximize the SDR at each resolution cell, i.e., when
\begin{equation}
\begin{cases}
 \bm C =\bm O_{MN, MN}\\
 P_r = P_{r,\text{max}}\\
 \ifCLASSOPTIONtwocolumn
 \bm w_{n,j} = \left(P_{r,\text{max}} \sum_{i=0}^{N-1} \sigma^2_{\gamma,i,j} \bm q_i \bm q_i^H + P_u\bm I_N \right)^{-1} \bm q_n, \\
 \hfill (n,j)\in\mathcal X.
 \else
 \bm w_{n,j} = \left(P_{r,\text{max}} \sum_{i=0}^{N-1} \sigma^2_{\gamma,i} \bm q_i \bm q_i^H + P_u\bm I_N \right)^{-1} \bm q_n, \quad (n,j)\in\mathcal X.
 \fi
\end{cases} \label{starting_point}
\end{equation}
From~\eqref{SDR_def}, this results in the following condition on $P_{r,\text{max}}$ and $\{\rho_{n,j}\}_{(n,j)\in\mathcal X}$ 
\ifCLASSOPTIONtwocolumn
\begin{multline}
 \sigma^2_{g,n,j} P_{r,\text{max}} \bm q_n^H \left(P_{r,\text{max}} \sum_{i=0}^{N-1} \sigma^2_{\gamma,i,j} \bm q_i \bm q_i^H + P_u\bm I_N \right)^{-1} \\
 \times \bm q_n \geq \rho_{n,j}, \quad \forall (n,j)\in\mathcal X. \label{rho_max}
\end{multline}
\else
\begin{equation}
 \sigma^2_{g,n,j} P_{r,\text{max}} \bm q_n^H \left(P_{r,\text{max}} \sum_{i=0}^{N-1} \sigma^2_{\gamma,i,j} \bm q_i \bm q_i^H + P_u\bm I_N \right)^{-1} \bm q_n \geq \rho_{n,j}, \quad \forall (n,j)\in\mathcal X.\label{rho_max}
\end{equation}
\fi
which will be assumed to be satisfied in the rest of the manuscript: should not it be satisfied, coexistence would not be possible.

Problem~\eqref{opt_prob} appears to be quite complex, so that we resort to the block coordinate ascent method~\cite{Bertsekas_1999}, also known as nonlinear Gauss-Seidel method or as alternating maximization: starting from a feasible point,\footnote{A feasible point is, clearly, that in~\eqref{starting_point}.} the objective function is maximized with respect to each of the ``block coordinate'' variables, taken in cyclic order, while keeping the other ones fixed at their previous values. If all maximizations are optimally solved (or, at least, the objective function is non-decreasing in successive maximizations), the algorithm converges. However, since the problem is not convex, and the feasible set cannot be expressed as the Cartesian product of closed convex sets, there is no guarantee that a global maximum is reached. In our setting, the natural block coordinate variables are $\big\{\{ \bm w_{n,j}\}_{(n,j)\in\mathcal X}, P_r, \bm C\big\}$, and we are faced with three reduced complexity sub-problems: radar receive filters optimization, radar transmit power optimization, and communication codebook optimization. These problems are solved in Secs.~\ref{radar_w_sol},~\ref {radar_Pr_sol}, and~\ref{opt_C_sol}, while, in Sec.~\ref {sub-opt_C_sol} a simple, sub-optimum solution to the communication codebook optimization is presented. Finally, the complete algorithm is reported in Sec.~\ref{algorithm_sec}, along with a discussion on its computational complexity.

\subsection{Radar receive filters optimization}\label{radar_w_sol}

The problem to be solved here is
\begin{equation}
 \begin{aligned}
 \max_{ \{ \bm w_{n,j}\}_{(n,j)\in\mathcal X} \in \mathbb C^N } & \quad R ( \bm C, P_r )\\
 \text{s.t.} & \quad \text{SDR}_{n,j} (\bm C, P_r, \bm w_{n,j}) \geq \rho_{n,j}, \; (n,j) \in \mathcal X.
 \end{aligned} \label{sub-prob_w}
\end{equation}
Since the objective function is independent of the radar filters, we select $\{\bm w_{n,j}\}_{(i,j)\in\mathcal X}$ so as to maximize the SDR in each resolution cell (and, therefore, guarantee the largest feasible set). Now, letting
\ifCLASSOPTIONtwocolumn
\begin{align}
 \bm D_j&=P_r\sum_{i=0}^{N-1} \sigma^2_{\gamma,i,j} \bm q_i \bm q_i^H +\sum_{i=0}^{N-1} \sum_{m=1}^M \sum_{m'=1}^M \sigma^2_{\beta,m,m',i,j}\notag\\
 & \quad \times (\bm A_{m,i} \bm C \bm A_{m',i}^T + \bm B_{m,i} \bm C \bm B_{m',i}^T) +P_u\bm I_N 
\end{align}
\else
\begin{equation}
 \bm D_j=P_r\sum_{i=0}^{N-1} \sigma^2_{\gamma,i,j} \bm q_i \bm q_i^H +\sum_{i=0}^{N-1} \sum_{m=1}^M \sum_{m'=1}^M \sigma^2_{\beta,m,m',i,j} (\bm A_{m,i} \bm C \bm A_{m',i}^T + \bm B_{m,i} \bm C \bm B_{m',i}^T) +P_u\bm I_N 
\end{equation}
\fi
the SDR in~\eqref{SDR_def} can be written as
\begin{equation}
 \text{SDR}_{n,j}(\bm C,P_r, \bm w_{n,j}) = \frac{\bm w_{n,j}^H \left(P_r \sigma^2_{g,n,j} \bm q_n \bm q_n^H\right) \bm w_{n,j}}{\bm w_{n,j}^H \bm D_j \bm w_{n,j}}
\end{equation}
which is a generalized Rayleigh quotient. Therefore
\begin{equation}
 \max_{\bm w_{n,j}\in\mathbb C^N} \text{SDR}_{n,j}(\bm C,P_r, \bm w_{n,j}) =P_r \sigma^2_{g,n,j}\bm q_n^H \bm D_j^{-1} \bm q_n
\end{equation}
and the maximum is achieved for
\ifCLASSOPTIONtwocolumn
 \begin{align}
 \bm w_{n,j}&\propto \bm D_j^{-1} \bm q_n\notag\\
 &=\left(P_r\sum_{i=0}^{N-1} \sigma^2_{\gamma,i,j} \bm q_i \bm q_i^H + \sum_{i=0}^{N-1} \sum_{m=1}^M \sum_{m'=1}^M \sigma^2_{\beta,m,m',i,j} \right. \notag\\
 & \quad \left.\vphantom{\sum_{i=0}^{N-1}} \times \big( \bm A_{m,i} \bm C \bm A_{m',i}^T+ \bm B_{m,i} \bm C \bm B_{m',i}^T \big) +P_u\bm I_N\right)^{-1} \bm q_n \label{w_update}
 \end{align}
\else
 \begin{align}
 \bm w_{n,j}\propto \bm D_j^{-1} \bm q_n &= \left(P_r\sum_{i=0}^{N-1} \sigma^2_{\gamma,i,j} \bm q_i \bm q_i^H \right. \notag\\
 &\quad \left.  + \sum_{i=0}^{N-1} \sum_{m=1}^M \sum_{m'=1}^M \sigma^2_{\beta,m,m',i,j} \big( \bm A_{m,i} \bm C \bm A_{m',i}^T + \bm B_{m,i} \bm C \bm B_{m',i}^T \big) +P_u\bm I_N\right)^{-1} \bm q_n\label{w_update}
 \end{align}
\fi
for $(i,j)\in\mathcal X$.

\subsection{Radar transmit power optimization}\label{radar_Pr_sol}

The problem to be solved here is
\begin{equation}
 \begin{aligned}
 \max_{P_r\in\mathbb R} & \quad R ( \bm C, P_r )\\
 \text{s.t.} & \quad \text{SDR}_{n,j} (\bm C, P_r, \bm w_{n,j}) \geq \rho_{n,j}, \; (n,j) \in \mathcal X\\
 & \quad 0\leq P_r \leq P_{r,\text{max}}.
 \end{aligned} \label{sub-prob_Pr}
\end{equation}
Since, from~\eqref{R_expr}, the objective function is strictly decreasing with the radar transmit power, we must select the smallest value of $P_r$ satisfying the SDR constraints. From~\eqref{SDR_def}, constraint $(i,j)$ can be rewritten as
\ifCLASSOPTIONtwocolumn
 \begin{align}
 P_r & \geq \rho_{n,j} \left( \sum_{i=0}^{N-1} \sum_{m=1}^M \sum_{m'=1}^M \sigma^2_{\beta,m,m',i,j} \bm w_{n,j}^H\right. \notag\\
 &\quad\times \left. \vphantom{\sum_{i=0}^{N-1}} \big( \bm A_{m,i} \bm C \bm A_{m',i}^T + \bm B_{m,i} \bm C \bm B_{m',i}^T \big) \bm w_{n,j} +P_u\Vert \bm w_{n,j}\Vert|^2 \right) \notag\\
 &\quad \times \left( \sigma^2_{g,n,j} |\bm w_{n,j}^H \bm q_n|^2 - \rho_{n,j} \sum_{i=0}^{N-1} \sigma^2_{\gamma,i,j} |\bm w_{n,j}^H \bm q_i|^2\right)^{-1}.
 \end{align}
\else
 \begin{equation}
P_r \geq \frac{\rho_{n,j} \left( \sum_{i=0}^{N-1} \sum_{m=1}^M \sum_{m'=1}^M \sigma^2_{\beta,m,m',i,j} \bm w_{n,j}^H \big( \bm A_{m,i} \bm C \bm A_{m',i}^T + \bm B_{m,i} \bm C \bm B_{m',i}^T \big)\bm w_{n,j} +P_u \Vert \bm w_{n,j}\Vert^2 \right) } {\sigma^2_{g,n,j} |\bm w_{n,j}^H \bm q_n|^2 - \rho_{n,j} \sum_{i=0}^{N-1} \sigma^2_{\gamma,i,j} |\bm w_{n,j}^H \bm q_i|^2}.
 \end{equation}
\fi
and, therefore,
\ifCLASSOPTIONtwocolumn
 \begin{align}
 P_r& = \max_{(n,j)\in\mathcal X} \rho_{n,j} \left( \sum_{i=0}^{N-1} \sum_{m=1}^M\sum_{m'=1}^M \sigma^2_{\beta,m,m',i,j} \bm w_{n,j}^H \right.\notag\\
 & \quad \times \big( \bm A_{m,i} \bm C \bm A_{m',i}^T+ \bm B_{m,i} \bm C \bm B_{m',i}^T \big) \bm w_{n,j} +P_u \Vert \bm w_{n,j}\Vert^2 \Bigg)\notag\\
 &\quad \times \left( \sigma^2_{g,n,j} |\bm w_{n,j}^H \bm q_n|^2 - \rho_{n,j} \sum_{i=0}^{N-1} \sigma^2_{\gamma,i,j} |\bm w_{n,j}^H \bm q_i|^2\right)^{-1}.\label{Pr_update}
 \end{align}
\else
 \begin{equation}
 P_r= \max_{(n,j)\in\mathcal X} \frac{\rho_{n,j} \left( \sum_{i=0}^{N-1} \sum_{m=1}^M \sum_{m'=1}^M \sigma^2_{\beta,m,m',i,j} \bm w_{n,j}^H \big( \bm A_{m,i} \bm C \bm A_{m',i}^T + \bm B_{m,i} \bm C \bm B_{m',i}^T \big)\bm w_{n,j} +P_u \Vert \bm w_{n,j}\Vert^2 \right) } {\sigma^2_{g,n,j} |\bm w_{n,j}^H \bm q_n|^2 - \rho_{n,j} \sum_{i=0}^{N-1} \sigma^2_{\gamma,i,j} |\bm w_{n,j}^H \bm q_i|^2}.\label{Pr_update}
 \end{equation}
\fi

\subsection{Communication codebook optimization} \label{opt_C_sol}

For the reader's sake, we first introduce some variables and notations that will allow rewriting the problem in a compact form. If we define
\ifCLASSOPTIONtwocolumn
\begin{align}
 \bm F &= \left(P_r \sum_{i=0}^{N-1} \bm \Sigma_{\alpha,i}\otimes( \bm q_i\bm q_i^H) +P_v \bm I_{KN} \right)^{-\frac12} \notag\\
 &\quad \times (\bm H \otimes \bm I_N) \in \mathbb C^{KN\times MN} \label{mat_F}
\end{align}
\else
 \begin{equation}
  \bm F = \left(P_r \sum_{i=0}^{N-1} \bm \Sigma_{\alpha,i}\otimes( \bm q_i\bm q_i^H) +P_v \bm I_{KN} \right)^{-\frac12} (\bm H \otimes \bm I_N) \in \mathbb C^{KN\times MN} \label{mat_F}
 \end{equation}
\fi
then, the objective function in Problem~\eqref{opt_prob} can be written as $\frac{1}{N} \log \det \left( \bm I_{KN} + \bm F \bm C \bm F^H \right)$. As to the SDR constraints, from~\eqref{SDR_def}, they can also be written as
\ifCLASSOPTIONtwocolumn
\begin{multline}
\trace \left\{ \bm C \sum_{i=0}^{N-1} \sum_{m=1}^M \sum_{m'=1}^M \sigma^2_{\beta,m,m',i,j} (\bm A_{m',i}^T \bm w_{n,j} \bm w_{n,j}^H \bm A_{m,i}\right.\\
\left. \vphantom{\sum_{i=0}^{N-1}} + \bm B_{m',i}^T \bm w_{n,j} \bm w_{n,j}^H\bm B_{m,i} ) \right\} \leq \frac{P_r \sigma^2_{g,n,j}}{\rho_{n,j}} |\bm w_{n,j}^H\bm q_n|^2\\
-P_r\sum_{i=0}^{N-1} \sigma^2_{\gamma,i,j} |\bm w_{i,j}^H\bm q_i|^2 - P_u \Vert \bm w_{n,j}\Vert^2.
\end{multline}
\else
\begin{multline}
\trace \left\{ \bm C \sum_{i=0}^{N-1} \sum_{m=1}^M \sum_{m'=1}^M \sigma^2_{\beta,m,m',i,j} (\bm A_{m',i}^T \bm w_{n,j} \bm w_{n,j}^H \bm A_{m,i} + \bm B_{m',i}^T \bm w_{n,j} \bm w_{n,j}^H\bm B_{m,i} ) \right\} \leq \frac{P_r \sigma^2_{g,n,j}}{\rho_{n,j}} |\bm w_{n,j}^H\bm q_n|^2\\
-P_r\sum_{i=0}^{N-1} \sigma^2_{\gamma,i,j} |\bm w_{i,j}^H\bm q_i|^2 - P_u \Vert \bm w_{n,j}\Vert^2.
\end{multline}
\fi
Therefore, if we let
\ifCLASSOPTIONtwocolumn
\begin{align}
 \bm E_{f(n,j)} & = \sum_{i=0}^{N-1} \sum_{m=1}^M \sum_{m'=1}^M \sigma^2_{\beta,m,m',i,j} (\bm A_{m',i}^T \bm w_{n,j} \bm w_{n,j}^H \bm A_{m,i} \notag \\
& \quad + \bm B_{m',i}^T \bm w_{n,j} \bm w_{n,j}^H\bm B_{m,i} )\in \mathbb C^{MN\times MN} , \quad (n,j)\in\mathcal X\\
 a_{f(n,j)} &= \frac{P_r \sigma^2_{g,n,j}}{\rho_{n,j}} |\bm w_{n,j}^H\bm q_n|^2 - P_r \sum_{i=0}^{N-1} \sigma^2_{\gamma,i,j} |\bm w_{n,j}^H \bm q_i|^2\notag \\
&\quad -P_u\Vert \bm w_{n,j}\Vert^2 \geq0, \quad (n,j)\in\mathcal X
\end{align}
\else
\begin{align}
 \bm E_{f(n,j)} & = \sum_{i=0}^{N-1} \sum_{m=1}^M \sum_{m'=1}^M \sigma^2_{\beta,m,m',i,j} (\bm A_{m',i}^T \bm w_{n,j} \bm w_{n,j}^H \bm A_{m,i}+ \bm B_{m',i}^T \bm w_{n,j} \bm w_{n,j}^H\bm B_{m,i} )\in \mathbb C^{MN\times MN} , \quad (n,j)\in\mathcal X\\
 a_{f(n,j)} &= \frac{P_r \sigma^2_{g,n,j}}{\rho_{n,j}} |\bm w_{n,j}^H\bm q_n|^2 - P_r \sum_{i=0}^{N-1} \sigma^2_{\gamma,i,j} |\bm w_{n,j}^H \bm q_i|^2-P_u\Vert \bm w_{n,j}\Vert^2 \geq0, \quad (n,j)\in\mathcal X
\end{align}
\fi
where $f$ is a one-to-one mapping from $\mathcal X$ to $\big\{1,\ldots,\card(\mathcal X) \big\}$, the SDR constrains can be rewritten as $\trace \{ \bm E_\ell \bm C \} \leq a_\ell$, $\ell=1,\ldots,\card(\mathcal X)$. With this notation, the problem to be solved here becomes
\begin{equation}
 \begin{aligned}
 \max_{\bm C \in \mathbb C^{MN\times MN}} & \quad \log \det \left( \bm I_{KN} + \bm F \bm C \bm F^H \right) \\
 \text{s.t} & \quad \bm C \text{ Hermitian positive semi-definite}\\
 & \quad \trace \{\bm E_\ell \bm C \} \leq a_\ell, \quad \ell=1,\ldots,U
 \end{aligned} \label{sub-prob_C}
\end{equation}
where $U=\card(\mathcal X)+1$, and the power constraint is handled by $\bm E_U =\bm I_{MN}$ and $a_U = N P_{c,\text{max}}$.

Problem~\eqref{sub-prob_C} is a determinant maximization with linear matrix inequalities~\cite{Vandenberghe_1998}: it is a convex optimization problem that can be readily solved by standard interior-point methods, at least when the number of variables is modest. When $N$ is large, first order methods are instead preferable, so as to reduce complexity and allow a real-time implementation. In this case, we adopt a (sub)gradient method for constrained optimization\footnote{We do not consider the projected gradient algorithm, since it would be computationally more demanding: it would require, at each iteration, a projection on the constraint set, which is itself an optimization problem, that, following~\cite{Boyd_2005}, can be solved with another projected gradient algorithm.}~\cite{Boyd_2014}, and we handle the positive semidefinite constraint on $\bm C$ by introducing the auxiliary variable $\bm X \in \mathbb C^{MN \times MN}$ such that $\bm C = \bm X \bm X^H$. Problem~\eqref{sub-prob_C} then becomes
\begin{equation}
 \begin{aligned}
 \max_{\bm X \in \mathbb C^{MN\times MN}} & \quad \log \det \left( \bm I_{KN} + \bm F \bm X \bm X^H \bm F^H \right) \\
 \text{s.t} & \quad \trace \{\bm X^H \bm E_\ell \bm X \} \leq a_\ell, \quad \ell=1,\ldots,U
 \end{aligned} \label{X_opt_prob}
\end{equation}
and the algorithm takes the simple form
\begin{equation}
 \bm X_{k+1} = \bm X_k + \alpha_k G \big(\bm X_k\big) \label{update_X_eq}
\end{equation}
where $\alpha_k$ is the step-size at the $k$-th iteration, and $G \big(\bm X_k\big)$ is the gradient of the objective function, if $\bm X_k$ is feasible, and the opposite of the gradient of any violated constraint, otherwise. Precisely, if $\trace \left\{\bm X_k^H \bm E_\ell \bm X_k \right\} \leq a_\ell$, for all $\ell\in\{1,\ldots, U\}$, then
\begin{equation}
 G (\bm X_k)= \bm F^H \bm F \bm X_k \big( \bm I_{MN} + \bm X_k^H \bm F^H \bm F \bm X_k \big)^{-1} \label{gradient_obj}
\end{equation}
where we have exploited~\cite[Eq.~(23)]{Palomar_1996} to evaluate the gradient of the mutual information. If, instead, $\trace \left\{\bm X_k^H \bm E_\ell \bm X_k \right\} > a_\ell$ for some $\ell\in\{1,\ldots,U\}$, then
\begin{equation}
 G (\bm X_k)= - \sum_{\substack{\ell\in\{1,\ldots,U\}:\\ \trace \{\bm X_k^H \bm E_\ell \bm X_k \} > a_\ell}} \bm E_\ell \bm X_k
\end{equation}
where we have included the gradient of all violated constraint, as in~\cite{Gerards_2008}. Since the objective function can decrease over consecutive iterations ($G (\bm X_k)$ may not be a descent direction and/or the step-size $\alpha_k$ can can be too large), it is common to keep track of the best point found so far, i.e., the one with largest function value:
\begin{equation}
 \bm X_{k,\text{best}} = \argmax_{\substack{ \{\bm X_i\}_{i=1}^k: \\ \trace \{\bm X_i^H \bm E_\ell \bm X_i\} \leq a_\ell \; \forall \ell}} \log \det \left( \bm I_{KN} + \bm F \bm X_i \bm X_i^H \bm F^H \right). \label{X_best_track}
\end{equation}
Notice that, unlike~\eqref{sub-prob_C}, Problem~\eqref{X_opt_prob} is not a convex optimization, and the gradient method in~\eqref{update_X_eq} does not guarantee that a global maximum is achieved. Nevertheless, since the objective function evaluated at $\bm X_{k,\text{best}}$ is non-decreasing with $k$, the gradient method converges and returns an updated covariance matrix $\bm C=\bm X_{k,\text{best}}\bm X_{k,\text{best}}^H$ resulting in a mutual information greater than or equal to the one obtained at the previous iteration of the block-coordinate descent method.

\subsection{A sub-optimum communication codebook}\label{sub-opt_C_sol}

Problem~\eqref{sub-prob_C} in the previous section can be simplified if we sub-optimally fix the eigenvectors of the matrix $\bm C$ and maximize over its eigenvalues only. Let $\bm U \bm \Xi \bm V^H$ be the singular value decomposition (SVD) of $\bm F$, where $\bm U\in\mathbb C^{KN\times KN}$ and $\bm V \in \mathbb C^{MN \times MN}$ are unitary matrices, and $\bm \Xi \in \mathbb R^{KN \times KM}$ is a diagonal matrix with non-negative entries (sorted in decreasing order) on the principal diagonal. Let also $\Delta\leq N\min\{K,M\}$ be the number of non-zero singular values. From Hadamard's inequality, the objective function of Problem~\eqref{sub-prob_C} is maximized when $\bm F \bm C \bm F^H$ is diagonal, so we (sub-optimally) set $\bm C$ as\footnote{Since the power outside the column span of $\bm V$ does not increase the objective function, the remaining $MN-\Delta$ eigenvalues of $\bm C$ can be set equal to zero. The choice in~\eqref{C_structure} would be optimum, if the SDR constraints are not present~\cite{Cover_Thomas_2006}.}
\begin{equation}
 \bm C = \bm V \diag (p_1, \ldots, p_{\Delta}, 0,\ldots,0) \bm V^H . \label{C_structure}
\end{equation}

With this choice, Problem~\eqref{sub-prob_C} simplifies to
\begin{equation}
 \begin{aligned}
 \max_{p_1, \ldots, p_{\Delta} \in \mathbb R} & \quad \sum_{i=1}^{\Delta} \log \left( 1+\frac{p_i}{\sigma^2_i} \right) \\
 \text{s.t} & \quad p_i\geq 0, \quad i=1,\ldots,{\Delta}\\
 & \quad \sum_{i=1}^{\Delta} e_{\ell,i} p_i \leq a_\ell, \quad \ell=1,\ldots,U
 \end{aligned}\label{C_opt_prob_simpl}
\end{equation}
where $e_{\ell,i}= [\bm V^H \bm E_\ell \bm V]_{i,i}\geq0$, and $\sigma^2_i=1/\Xi_{i,i}^2$. Notice that $p_i$ is the power transmitted along the $i$-th eigenmode of the MIMO channel represented by $\bm F$, and $\sigma^2_i$ is the corresponding disturbance (noise plus interference from the radar); furthermore, since $\bm E_U = \bm I_{MN}$ and $a_U=P_{c,\text{max}}$, the last constrain in~\eqref{C_opt_prob_simpl} is just the power constraint $\sum_{i=1}^{\Delta} p_i \leq P_{c,\text{max}}$.

To solve Problem~\eqref{C_opt_prob_simpl}, we introduce the Lagrange multipliers $\lambda_i$, associated with the constraints $p_i\geq0$, for $i=1,\ldots,\Delta$, and $\mu_\ell$, associated with the constraints $\sum_{i=1}^{\Delta} e_{\ell,i} p_i \leq a_\ell$, for $\ell=1,\ldots,U$; the Lagrangian is, therefore,
\ifCLASSOPTIONtwocolumn
 \begin{align}
 \mathcal L (\bm p, \bm \lambda, \bm \mu) &= -\sum_{i=1}^{\Delta} \log \left( 1+\frac{p_i}{\sigma^2_i}\right) - \sum_{i=1}^{\Delta} \lambda_i p_i \notag\\
 & \quad +\sum_{\ell=0}^U\mu_\ell \left(\sum_{i=1}^{\Delta} e_{\ell,i} p_i -a_\ell\right) \notag\\
 &=\sum_{i=1}^{\Delta}\left[(\bm e_i^T \bm \mu -\lambda_i)p_i - \log\left( 1+\frac{p_i}{\sigma^2_i}\right) \right]- \bm a^T \bm \mu
 \end{align}
\else
 \begin{align}
 \mathcal L (\bm p, \bm \lambda, \bm \mu) &= -\sum_{i=1}^{\Delta} \log \left( 1+\frac{p_i}{\sigma^2_i}\right) - \sum_{i=1}^{\Delta} \lambda_i p_i +\sum_{\ell=0}^U\mu_\ell \left(\sum_{i=1}^{\Delta} e_{\ell,i} p_i -a_\ell\right) \notag\\
 &=\sum_{i=1}^{\Delta}\left[(\bm e_i^T \bm \mu -\lambda_i)p_i - \log\left( 1+\frac{p_i}{\sigma^2_i}\right) \right]- \bm a^T \bm \mu
 \end{align}
\fi
where $\bm p=(p_1 \,\cdots \, p_{\Delta})^T$, $\bm \lambda=(\lambda_1 \,\cdots \, \lambda_{\Delta})^T$, $\bm \mu=(\mu_1 \,\cdots \, \mu_U)^T$, and $\bm e_i=(e_{1,i} \,\cdots \, e_{U,i})^T$, $i=1,\ldots,\Delta$. The dual function is the infimum of the Lagrangian over $\bm p$
\begin{equation}
 g(\bm \lambda, \bm \mu) = \inf_{\bm p\in\mathbb R^{\Delta}: p_i>-\sigma^2_i} \mathcal L (\bm p, \bm \lambda, \bm \mu)
\end{equation}
that yields
\begin{equation}
 p_i =\begin{cases}
 \frac{1}{\bm e_i^T \bm \mu-\lambda_i}- \sigma^2_i, & \text{if } \bm e_i^T \bm \mu-\lambda_i>0\\
 \infty, & \text{otherwise}
 \end{cases}\label{opt_p_1}
\end{equation}
for $i=1,\ldots, \Delta$, so that
\ifCLASSOPTIONtwocolumn
\begin{align}
 g(\bm \lambda, \bm \mu) &= \sum_{i=1}^{\Delta}\left[ \left(1-\sigma^2_i(\bm e_i^T \bm \mu -\lambda_i) + \log \big(\sigma^2_i(\bm e_i^T \bm \mu -\lambda_i) \big)\right) \vphantom{\times \mathbbm 1_{\{\bm e_i^T \bm \mu > \lambda_i\}} -\infty \mathbbm 1_{\{\bm e_i^T \bm \mu \leq \lambda_i\}} }\right.\notag\\
 & \quad \left. \vphantom{1-\sigma^2_i(\bm e_i^T \bm \mu -\lambda_i) + \log \big(\sigma^2_i(\bm e_i^T \bm \mu -\lambda_i) \big)}\times \mathbbm 1_{\{\bm e_i^T \bm \mu > \lambda_i\}} -\infty \mathbbm 1_{\{\bm e_i^T \bm \mu \leq \lambda_i\}} \right]- \bm a^T \bm \mu .
\end{align}
\else
\begin{equation}
 g(\bm \lambda, \bm \mu) = \sum_{i=1}^{\Delta}\left[ \left(1-\sigma^2_i(\bm e_i^T \bm \mu -\lambda_i) + \log \big(\sigma^2_i(\bm e_i^T \bm \mu -\lambda_i) \big)\right) \vphantom{\times \mathbbm 1_{\{\bm e_i^T \bm \mu > \lambda_i\}} -\infty \mathbbm 1_{\{\bm e_i^T \bm \mu \leq \lambda_i\}} }\mathbbm 1_{\{\bm e_i^T \bm \mu > \lambda_i\}} -\infty \mathbbm 1_{\{\bm e_i^T \bm \mu \leq \lambda_i\}} \right]- \bm a^T \bm \mu .
\end{equation}
\fi
The dual problem is, therefore,
\begin{equation}
\begin{aligned}
 \max_{\bm \lambda\in\mathbb R^{\Delta}, \bm \mu \in \mathbb R^U}& \quad g(\bm \lambda, \bm \mu) \\
 \text{s.t.} & \quad \lambda_i\geq 0, \quad i=1,\ldots, \Delta\\
 & \quad \mu_\ell\geq 0, \quad \ell=1,\ldots, U \label{dual_problem}
\end{aligned}
\end{equation}

Weak duality always holds for the dual problem in~\eqref{dual_problem}: if $\bm \lambda$ and $\bm \mu$ are dual feasible, i.e., $\lambda_i\geq0$ for all $i$ and $\mu_\ell\geq0$ for all $\ell$, then the dual objective is a lower bound to the optimal value of Problem~\eqref{C_opt_prob_simpl}. If Problem~\eqref{C_opt_prob_simpl} is strictly feasible, i.e., there exists a $\bm p$ satisfying the linear inequalities and such that $p_i>0$ for some $i$, then strong duality holds: there exists $\bm \lambda^*$ and $\bm \mu^*$ that are optimal for the dual problem in~\eqref{dual_problem} with dual objective equal to the optimal value of Problem~\eqref{C_opt_prob_simpl}. Moreover, we can recover the solution to Problem~\eqref{C_opt_prob_simpl} from the dual optimal variables using~\eqref{opt_p_1}. This follows from convexity of Problem~\eqref{C_opt_prob_simpl} and strict convexity of the Lagrangian with respect to the primal variable $\bm p$. Observe that the condition $p_i>0$ for some $i$ simply means that the communication system can satisfy the constraint of Problem~\eqref{C_opt_prob_simpl} with some $\bm C\neq \bm O_{MN,MN}$, i.e., that the original problem admits a solution where coexistence is allowed.

Looking at Problem~\eqref{dual_problem}, one notice that the maximization over $\bm \lambda$ can be carried out analytically, and
\ifCLASSOPTIONtwocolumn
 \begin{align}
 \max_{\bm \lambda\in\mathbb R^{\Delta}: \lambda_i\geq 0} & g(\bm \lambda, \bm \mu) \notag\\
 &= \sum_{i=1}^{\Delta}\left[ \left( 1-\sigma^2_i \bm e_i^T \bm \mu + \log \sigma^2_i \bm e_i^T \bm \mu \right) \mathbbm 1_{\{0<\sigma^2_i \bm e_i^T \bm \mu\leq 1\}} \right.\notag\\
 & \quad \vphantom{\sum_{i=1}^{\Delta}} \left. -\infty \mathbbm 1_{\{\sigma^2_i \bm e_i^T \bm \mu\leq 0\}}\right]- \bm a^T \bm \mu 
 \end{align}
\else
 \begin{equation}
 \max_{\bm \lambda\in\mathbb R^{\Delta}: \lambda_i\geq 0} g(\bm \lambda, \bm \mu) = \sum_{i=1}^{\Delta}\left[ \left( 1-\sigma^2_i \bm e_i^T \bm \mu + \log \sigma^2_i \bm e_i^T \bm \mu \right) \mathbbm 1_{\{0<\sigma^2_i \bm e_i^T \bm \mu\leq 1\}} -\infty \mathbbm 1_{\{\sigma^2_i \bm e_i^T \bm \mu\leq 0\}}\right]- \bm a^T \bm \mu
 \end{equation}
\fi
for $\lambda_i = \left(\bm e_i^T \bm \mu-1/\sigma^2_i\right)^+$, $i=1,\ldots,\Delta$. This leads to the simplified dual problem
\ifCLASSOPTIONtwocolumn
 \begin{equation}
 \begin{aligned}
 \max_{\bm \mu \in \mathbb R^U}& \quad \left\{\sum_{i=1}^{\Delta}\left[ \left( 1-\sigma^2_i \bm e_i^T \bm \mu + \log \sigma^2_i \bm e_i^T \bm \mu \right) \mathbbm 1_{\{0<\sigma^2_i \bm e_i^T \bm \mu\leq 1\}} \right. \right.\\
 & \quad \left. \vphantom{\sum_{i=1}^{\Delta}} \left. \quad -\infty \mathbbm 1_{\{\sigma^2_i \bm e_i^T \bm \mu\leq 0\}}\right]- \bm a^T \bm \mu \right\}\\
 \text{s.t.} & \quad \mu_\ell\geq 0, \quad \ell=1,\ldots, U
 \end{aligned}\label{dual_problem_simpl}
 \end{equation}
\else
\begin{equation}
 \begin{aligned}
 \max_{\bm \mu \in \mathbb R^U}& \quad \left\{\sum_{i=1}^{\Delta}\left[ \left( 1-\sigma^2_i \bm e_i^T \bm \mu + \log \sigma^2_i \bm e_i^T \bm \mu \right) \mathbbm 1_{\{0<\sigma^2_i \bm e_i^T \bm \mu\leq 1\}}  -\infty \mathbbm 1_{\{\sigma^2_i \bm e_i^T \bm \mu\leq 0\}}\right]- \bm a^T \bm \mu \right\}\\
 \text{s.t.} & \quad \mu_\ell\geq 0, \quad \ell=1,\ldots, U
 \end{aligned}\label{dual_problem_simpl}
 \end{equation}
\fi
and, from~\eqref{opt_p_1}, to the solution to the primal problem in~\eqref{C_opt_prob_simpl}
\begin{equation}
 p_i=\left(\frac{1}{\bm e_i^T \bm \mu}-\sigma^2_i\right)^+, \quad i=1,\ldots,\Delta. \label{opt_rho_2}
\end{equation}
Observe that, should the power constraint only be present, then $\bm \mu$ and $\bm e_i$ would be scalars, with $\bm e_i=1$, and the solution would reduce to water-filling~\cite{Cover_Thomas_2006}.

The simplified dual problem is clearly a convex optimization problem. The gradient of the objective function is
\begin{equation}
 \sum_{i=1}^{\Delta} \left(\frac{1}{\bm e_i^T \bm \mu}-\sigma^2_i\right)^+ \bm e_i -\bm a
\end{equation}
whose entries are just the residuals for the constraints in the primal problem with $\bm p$ as in~\eqref{opt_rho_2}.
Several methods can be used to solve the simplified dual problem. Here we adopt a very simple one: the projected (sub)gradient method. At each step, the algorithm selects the next point by moving towards the gradient direction (like the gradient method) and then projecting onto the feasible set; in the present case, the projection operation is just the positive part of each coordinate of the point. The iterative step of the algorithm is, therefore,
\begin{equation}
 \bm \mu_{k+1}= \left( \bm \mu_k + \alpha_k \left( \sum_{i=1}^{\Delta} \left(\frac{1}{\bm e_i^T \bm \mu_k}-1\right)^+ \bm e_i -\bm a \right)\right)^+ \label{mu_update}
 \end{equation}
where $\alpha_k$ is the step-size parameter at the $k$-th iteration. Many choices are available for the step-size~\cite{Shor_1985, Shor_1988}, and a simple one that guarantees convergence is $\alpha_k$ such that $\alpha_k\geq 0$, $\lim_{k\rightarrow \infty} \alpha_k=0$, and $\sum_{k=1}^\infty \alpha_k=\infty$ (e.g., a universal choice is $\alpha_k=1/k$).

\subsection{The complete algorithm} \label{algorithm_sec}

The block coordinate ascent method, used to find a sub-optimum solution to Problem~\eqref{opt_prob}, and the gradient method, used to solve the communication codebook optimization problem in Sec.~\ref{opt_C_sol}, are integrated in Algorithm~\ref{alg_1}. Loops end when a sufficiently small percentage increase of the objective function in two consecutive iterations is observed or when a specified maximum number of iterations is reached. The computational complexity of the inner loop of Algorithm~\ref{alg_1} is dominated by the matrix inversion in~\eqref{gradient_obj}, needed to update $\bm X$, and by the evaluation of the objective function in Problem~\eqref{X_opt_prob}, needed to keep track of the $\bm X_{k,\text{best}}$: the former has a cost of $\mathcal O(N^3M^3)$, while the latter has a cost of $\mathcal O(N^3K^3)$, so that the overall cost is $\mathcal O\big(N^3\max\{M^3,K^3\}\big)$. The computational complexity of the remaining operations in the outer loop is dictated by the radar filters and power update in~\eqref{w_update} and~\eqref{Pr_update}, and by the evaluation of matrices $\bm F$ in~\eqref{mat_F} and $\bm F^H \bm F$, needed in the inner loop: the first two operations have a cost of $\mathcal O(N^3J)$, the third has a cost of $\mathcal O\big(N^3K^2\max\{M,K\}\big)$, and the forth has a cost of $\mathcal O(N^3M^2K)$, so that the overall cost is $\mathcal O\big(N^3\max\{J,M^2K,K^3\}\big)$.

\begin{algorithm}[t] 
\caption{Sub-optimum solution to Problem~\eqref{opt_prob}}
\label{alg_1}
 \begin{algorithmic}
 \STATE chose $\{\bm w_{n,j}\}_{(n,j)\in\mathcal X}$, $P_r$, $\bm C$ satisfying the constraints
 \REPEAT
 \STATE update $\{\bm w_{n,j}\}_{(n,j)\in\mathcal X}$ with~\eqref{w_update}
 \STATE update $P_r$ with~\eqref{Pr_update}
 \STATE choose $\bm X \in \mathbb C^{MN\times MN}$: $\bm X \neq \bm O_{MN, MN}$
 \REPEAT
 \STATE update $\bm X$ with~\eqref{update_X_eq}
 \STATE keep track of $\bm X_{k,\text{best}}$ with~\eqref{X_best_track}
 \UNTIL{convergence}
 \STATE update $\bm C$ with $\bm C=\bm X_{k,\text{best}}\bm X_{k,\text{best}}^H$
 \UNTIL{convergence}
 \RETURN $\{\bm w_{n,j}\}_{(n,j)\in\mathcal X}$, $P_r$, $\bm C$
 \end{algorithmic}
\end{algorithm}

Algorithm~\ref{alg_2}, instead, uses the sub-optimum procedure described in Sec.~\ref{sub-opt_C_sol} to solve the communication codebook optimization problem. The computational complexity of the inner loop of Algorithm~\ref{alg_2} is now ruled by the update of $\bm \mu$ in~\eqref{mu_update}, that has a cost of $\mathcal O(U\Delta)$: since $\Delta\leq N\min\{M,K\}$ and $U\leq (N-L+1)J$, the cost is at most $\mathcal O\big(N^2 J \min\{M,K\}\big)$. This complexity is smaller than that of the operations in the inner loop of Algorithm~\ref{alg_2} whenever $U\Delta < N^3 \max \{M^3,K^3\}$, that is always verified if $J\leq N\max\{M^3/K,K^3/M\}$. The remaining operations in the outer loop of Algorithm~\ref{alg_2} have the same complexity as in Algorithm~\ref{alg_1}, i.e., $\mathcal O\big(N^3\max\{J,M^2K,K^3\}\big)$.

\begin{algorithm}[t]
\caption{Sub-optimum solution to Problem~\eqref{opt_prob}}
\label{alg_2}
 \begin{algorithmic}
 \STATE chose $\{\bm w_{n,j}\}_{(n,j)\in\mathcal X}$, $P_r$, $\bm C$ satisfying the constraints
 \REPEAT
 \STATE update $\{\bm w_{n,j}\}_{(n,j)\in\mathcal X}$ with~\eqref{w_update}
 \STATE update $P_r$ with~\eqref{Pr_update}
 \STATE compute $(\bm U, \bm \Xi, \bm V)$, the SVD of $\bm F$ in~\eqref{mat_F}
 \STATE chose $\bm \mu\in\mathbb R_+^U: \bm \mu \neq \bm 0$
 \REPEAT
 \STATE update $\bm \mu$ with~\eqref{mu_update}
 \UNTIL{convergence}
 \STATE compute $\{p_i\}_{i=1}^\Delta$ with~\eqref{opt_rho_2}
 \STATE update $\bm C$ with~\eqref{C_structure}
 \UNTIL{convergence}
 \RETURN $\{\bm w_{n,j}\}_{(n,j)\in\mathcal X}$, $P_r$, $\bm C$
 \end{algorithmic}
\end{algorithm}

\section{Numerical examples}\label{num_ex_sec}

We examine two co-existing systems operating at 2~GHz over a bandwidth of 1.5~MHz. The radar has a PRF equal to 15~kHz, so that $N=100$ and $T=0.\bar 6$~$\mu$s, and uses a Barker code of length $L=5$. The maximum peak power is 500~W, so that $P_{r,\text{max}}=25$~W, and the maximum non-ambiguous range is 10~km. At the receiver side, $J=3$ orthogonal beams are formed; the power spectral density (PSD) of the noise is $\sigma^2_u= 4\times 10^{-21}$~W/Hz, so that $P_u=F\sigma^2_u W=2.39\times 10^{-14}$~W, where a receiver noise figure $F=6$~dB has been assumed. We test the system performance for a specified signal level, this corresponding to a target with increasing radar cross section located at increasing distance, and we require the same minimum SDR, $\rho$, at all range and azimuth bins; in particular, we set $\sigma^2_{g,n,j}=4.8\times10^{-16}$, for all $(n,j)\in\mathcal X$, so that $\sigma^2_{g,n,j} NP_{r,\text{max}} /P_u= 17$~dB (this is the largest achievable signal-to-noise ratio). As to the clutter component, we set $\sigma^2_{\gamma,i,j}=4.8\times10^{-17}$, for all $(n,j)\in\mathcal X$, so that $\sigma^2_{\gamma,i,j} NP_{r,\text{max}} /P_u= 7$~dB (this is the largest possible clutter-to-noise ratio in each resolution cell).

The communication system is equipped with $M=2$ transmit antennas and $K=2$ receive antennas; the maximum average transmit power is $P_{c,\text{max}}=10$~mW, and the distance between transmitter and receiver is 100~m. The PSD of the noise is $\sigma^2_v= 4\times 10^{-21}$~W/Hz, and $P_v=F\sigma^2_v W=2.39\times 10^{-14}$~W, where $F=6$~dB is the receiver noise figure. The entries of the channel matrix $\bm H$, perfectly estimated at the receiver, are generated following a CCSG distribution with variance $\sigma^2_h=3\times 10^{-10}$, so that $\sigma^2_h P_{c,\text{max}} / P_v=21$~dB (this is the largest achievable signal-to-noise ratio).

As to the mutual interference between the two systems, we set $\bm \Sigma_{\alpha,i}=\sigma^2_{\alpha,i}\bm I_K$ and $\sigma^2_{\beta,m,m',i,j}=\sigma^2_{\beta,i,j} \mathbbm 1_{\{m=m'\}}$ (no line-of-sight component is present, and independent rays from the rich scattering environment arrive at the receivers). The coefficients $\sigma^2_{\alpha,i}$ and $\sigma^2_{\beta,i,j}$ are equal to either 0 or $\sigma^2$, and we test $\sigma^2=1.2\times10^{-13}$ (resulting in $\sigma^2_{g,n,j} P_{r,\text{max}}/(\sigma^2P_{c,\text{max}}) = 10$ dB and $\sigma^2_h P_{c,\text{max}}/(\sigma^2P_{r,\text{max}})=0$ dB) and $\sigma^2=1.2\times10^{-11}$ (resulting in $\sigma^2_{g,n,j} P_{r,\text{max}}/(\sigma^2P_{c,\text{max}}) = -10$ dB and $\sigma^2_h P_{c,\text{max}}/(\sigma^2P_{r,\text{max}})=-20$ dB). The fraction of non-zero entries in $\{\sigma^2_{\alpha,i}\}_{i=0}^{N-1}$ and $\{\sigma^2_{\beta,i,j}\}_{i=0}^{N-1}$, for $j=1,\ldots,J$, is denoted $\delta$, and their indexes are randomly generated. Different values of $\delta$ are tested, and the performance is evaluated through Monte Carlo simulations.

\begin{figure}[t]
 \centering
 \ifCLASSOPTIONtwocolumn
 \centerline{\includegraphics[width=1.05\columnwidth]{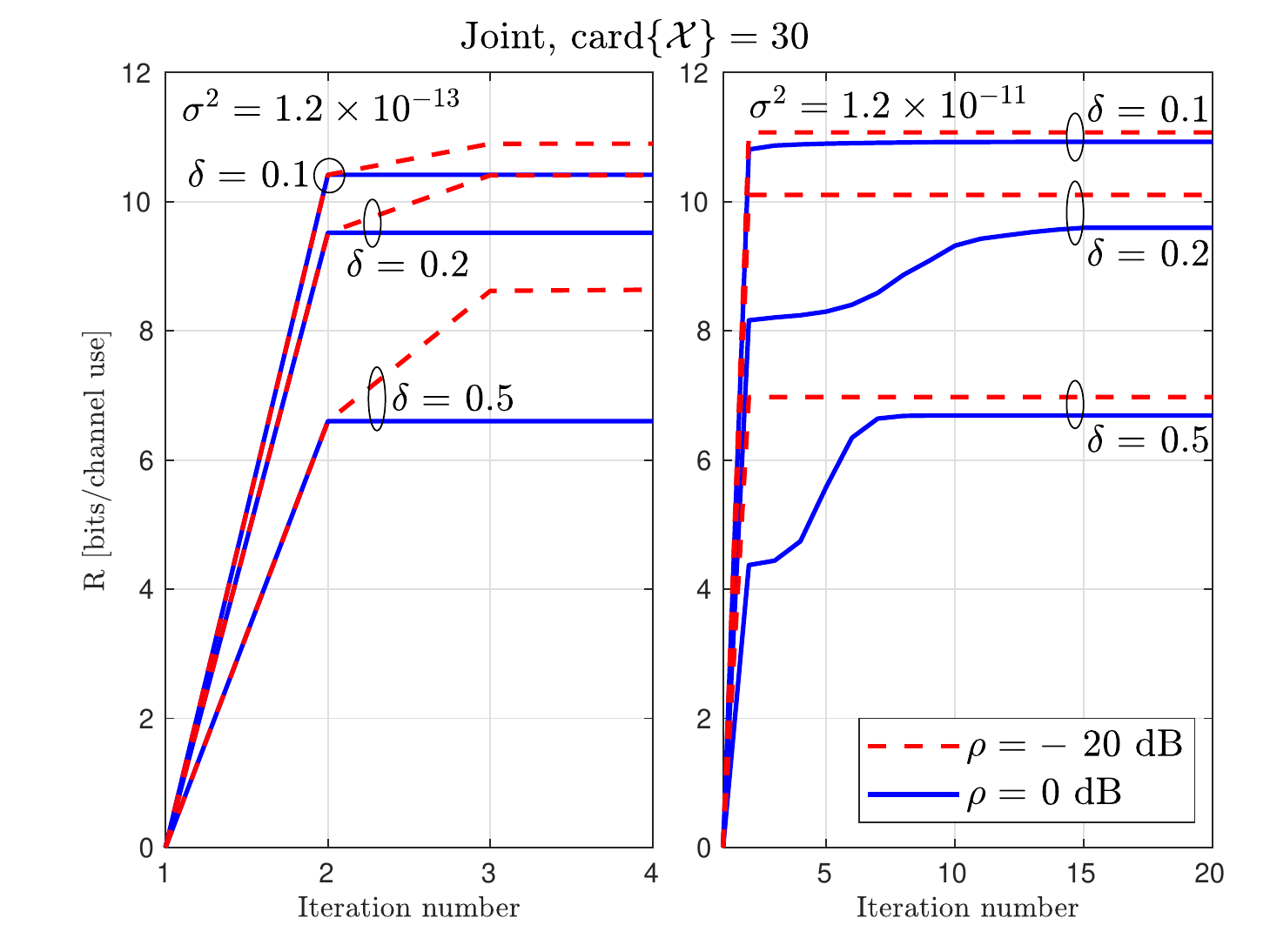}}
 \else
 \centerline{\includegraphics[width=0.75\columnwidth]{fig_04.pdf}}
 \fi
\caption{Mutual information versus the iteration number of the alternating maximization algorithm of the joint design strategy for different values of the minimum required SDR, $\rho$, intensity of the interference, $\sigma^2$, and density of the interference scatterers, $\delta$, when 30 resolution cells are protected and a joint design is undertaken.} \label{fig_04}
\end{figure}

\begin{figure}[t]
 \centering
 \ifCLASSOPTIONtwocolumn
 \centerline{\includegraphics[width=1.1\columnwidth]{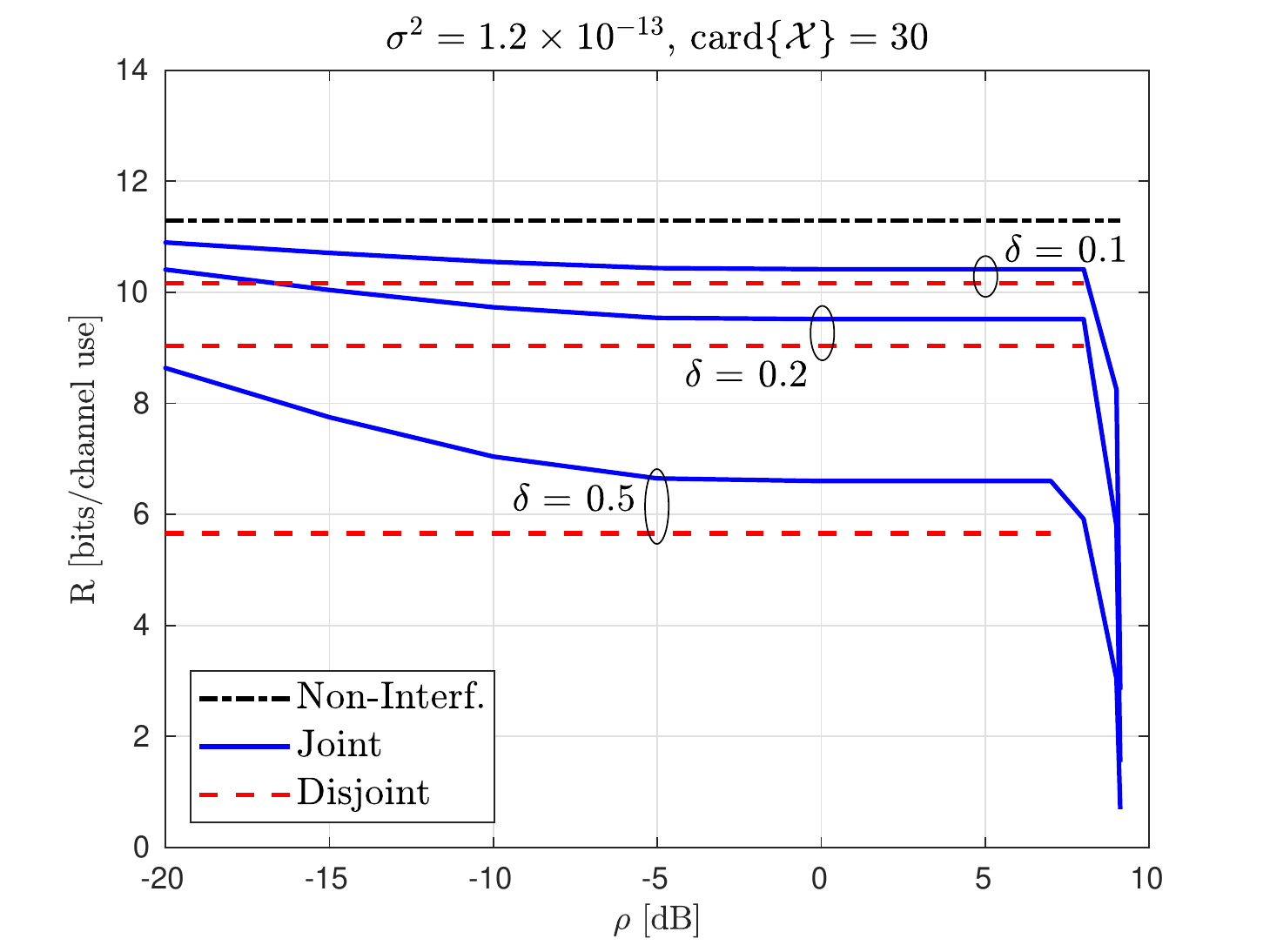}}
 \else
 \centerline{\includegraphics[width=0.75\columnwidth]{fig_05.pdf}}
 \fi
\caption{Mutual information versus the minimum required SDR for a joint and disjoint design, and for different values of the density of the interference scatterers, $\delta$, when the intensity of the interference is $\sigma^2=1.2\times 10^{-13}$, and 30 resolution cells are protected; for comparison purposes, the case of non-interfering systems is also included.} \label{fig_05}
\end{figure}

\begin{figure}[t]
 \centering
 \ifCLASSOPTIONtwocolumn
 \centerline{\includegraphics[width=1.1\columnwidth]{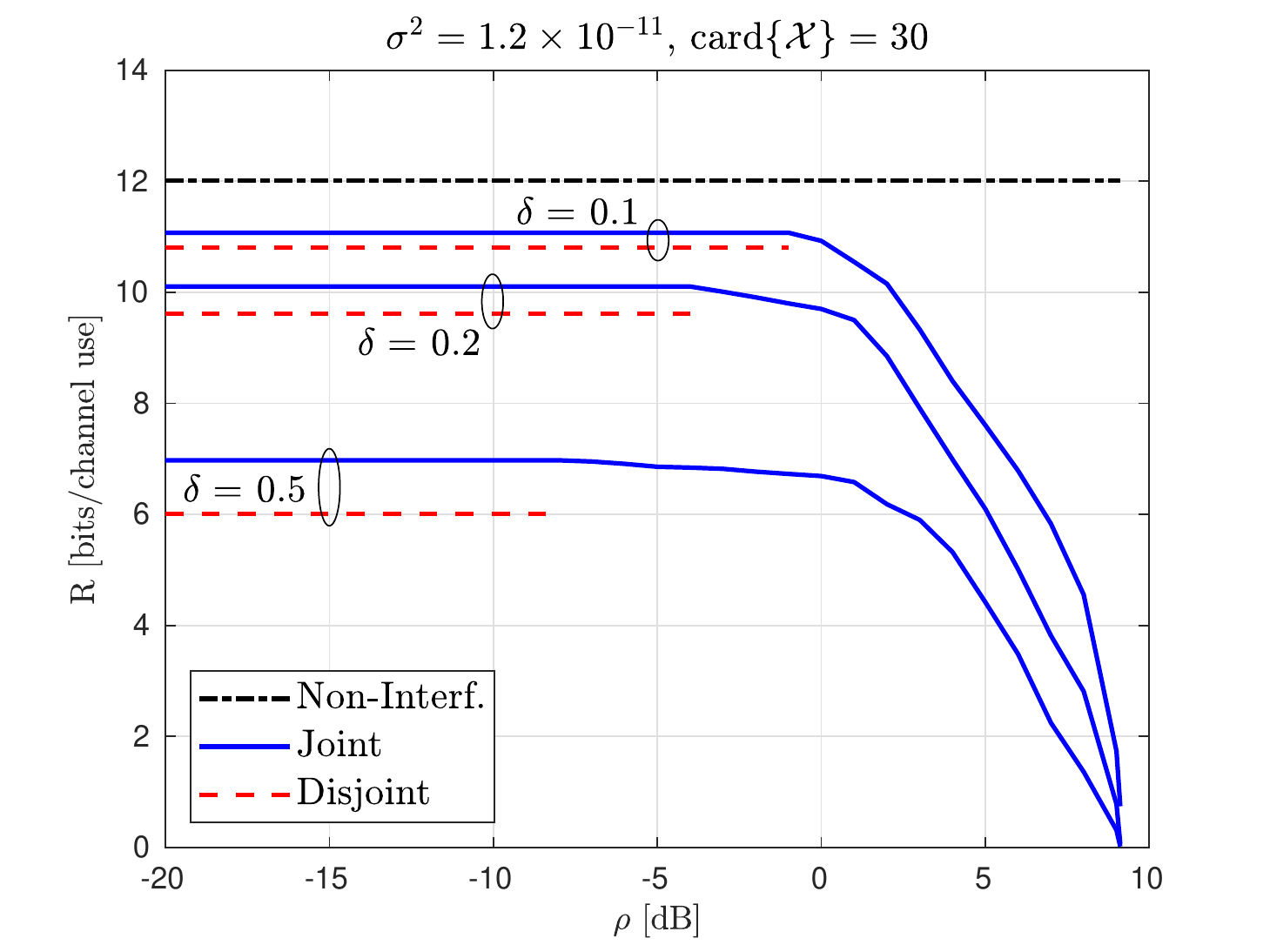}}
 \else
 \centerline{\includegraphics[width=0.75\columnwidth]{fig_06.pdf}}
 \fi
\caption{Mutual information versus the minimum required SDR for a joint and disjoint design, and for different values of the density of the interference scatterers, $\delta$, when the intensity of the interference is $\sigma^2=1.2\times 10^{-11}$, and 30 resolution cells are protected; for comparison purposes, the case of non-interfering systems is also included.} \label{fig_06}
\end{figure}

In Fig.~\ref{fig_04} the mutual information (in bits per channel use) of the proposed joint design is reported versus the iteration number of the alternating maximization algorithm for different values of the density of the interference scatterers, $\delta$, intensity of the interference, $\sigma^2$, and minimum required SDR at the radar, $\rho$, when the number of protected radar resolution cells is $\card\{\mathcal X\}=30$. It can be seen that the algorithm rapidly converges in all inspected cases, requiring a slightly larger number of iterations only when there is a strong interference from a dense scattering environment and a high performance is requested at the radar side.

Fig.~\ref{fig_05} shows the mutual information versus $\rho$ for different values of $\delta$, when $\sigma^2=1.2\times 10^{-13}$ and $\card\{\mathcal X\}=30$. Notice that not all the values of the SDR constraint are feasible, since noise and clutter are always present: in fact, from~\eqref{rho_max}, Problem~\eqref{opt_prob} admits a solution only if $\rho\leq 9.12$~dB. For comparison purposes, we also include the following two cases. In the first one, referred to as \emph{non-interfering} systems, Problem~\eqref{opt_prob} is solved when there is no mutual interference ($\delta = 0$ and/or $\sigma^2=0$). A solution at the radar side is
\begin{subequations}
\begin{align}
 P_r&=P_{r,\text{max}}\\
 \bm w_{n,j}&=\left(P_{r,\text{max}}\sum_{i=0}^{N-1} \sigma^2_{\gamma,i,j} \bm q_i \bm q_i^H +P_u\bm I_N\right)^{-1} \bm q_n, \; (n,j)\in\mathcal X
\end{align}\label{radar_only}%
\end{subequations}
while, at the communication side, $\bm C$ can be found through standard waterfilling over the channel $\bm r = (\bm H \otimes \bm I_N) \bm c+ \bm v$ (therefore, $P_c=P_{c,\text{max}}$). This gives the same value of mutual information independently of the constraint $\rho$ as long as $\rho \leq 9.12$~dB, and represents an upper bound to the system performance. In the second case, the previous solution is incorrectly used when the mutual interference is present; this corresponds to the case where each system independently maximizes its own performance measure (mutual information/minimum SDR) ignoring the presence of the other system, and is therefore referred to as \emph{disjoint design}. The curves corresponding to this case are, again, half-lines with zero slope, but the ending points are shifted towards smaller values of $\rho$ due to the incorrect assumption of absence of interference. By inspecting Fig.~\ref{fig_05}, we see that the mutual information of the proposed joint design decreases with $\rho$ and $\delta$, and achieves the upper bound of the non-interfering case when $\rho\rightarrow -\infty$. Notice that the joint design outperforms the disjoint one not only in the value of mutual information, where the gap becomes significant for high $\delta$'s and/or low $\rho$'s, but also in the achievable values of minimum SDR, where the gap is approximately 1 or 2 dB. The gain of the joint design is more evident when the mutual interference is stronger, as it can be seen from Fig.~\ref{fig_06}, that examines the case $\sigma^2=1.2\times 10^{-11}$. In this situation, the gap in terms of achievable SDR with respect to the disjoint design is significant, and amounts to as much as 10, 13, and 17~dB for $\delta=0.1$, 0.2, and 0.5, respectively.

\begin{figure}[t]
 \centering
 \ifCLASSOPTIONtwocolumn
 \centerline{\includegraphics[width=1.1\columnwidth]{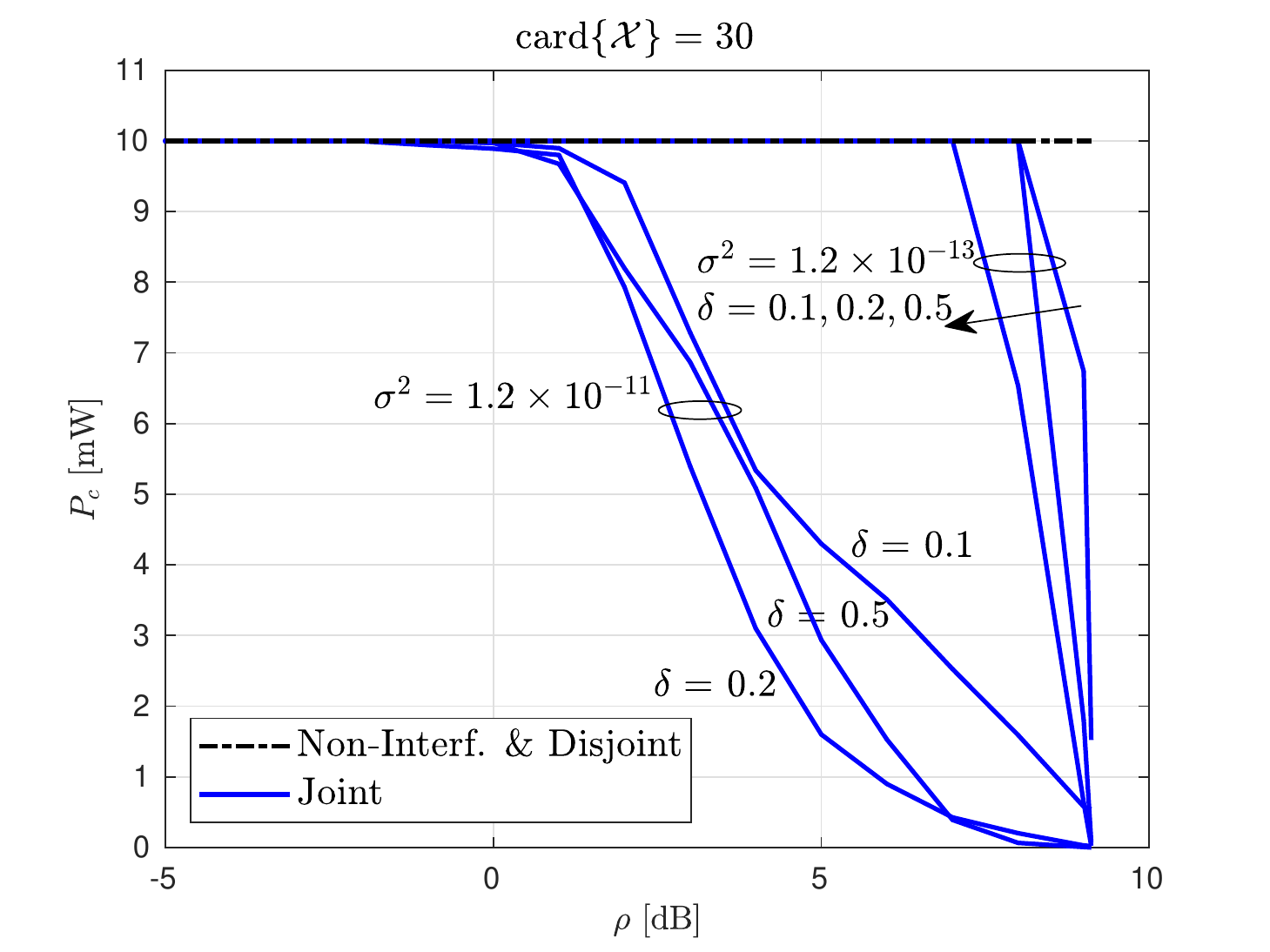}}
 \else
 \centerline{\includegraphics[width=0.75\columnwidth]{fig_07.pdf}}
 \fi
\caption{Communication transmit power versus the minimum required SDR for a joint and disjoint design, different values of the density of the interference scatterers, $\delta$, and intensity of the interference, $\sigma^2$, when 30 resolution cells are protected; for comparison purposes, the case of non-interfering systems is also included.} \label{fig_07}
\end{figure}

\begin{figure}[t]
 \centering
 \ifCLASSOPTIONtwocolumn
 \centerline{\includegraphics[width=1.1\columnwidth]{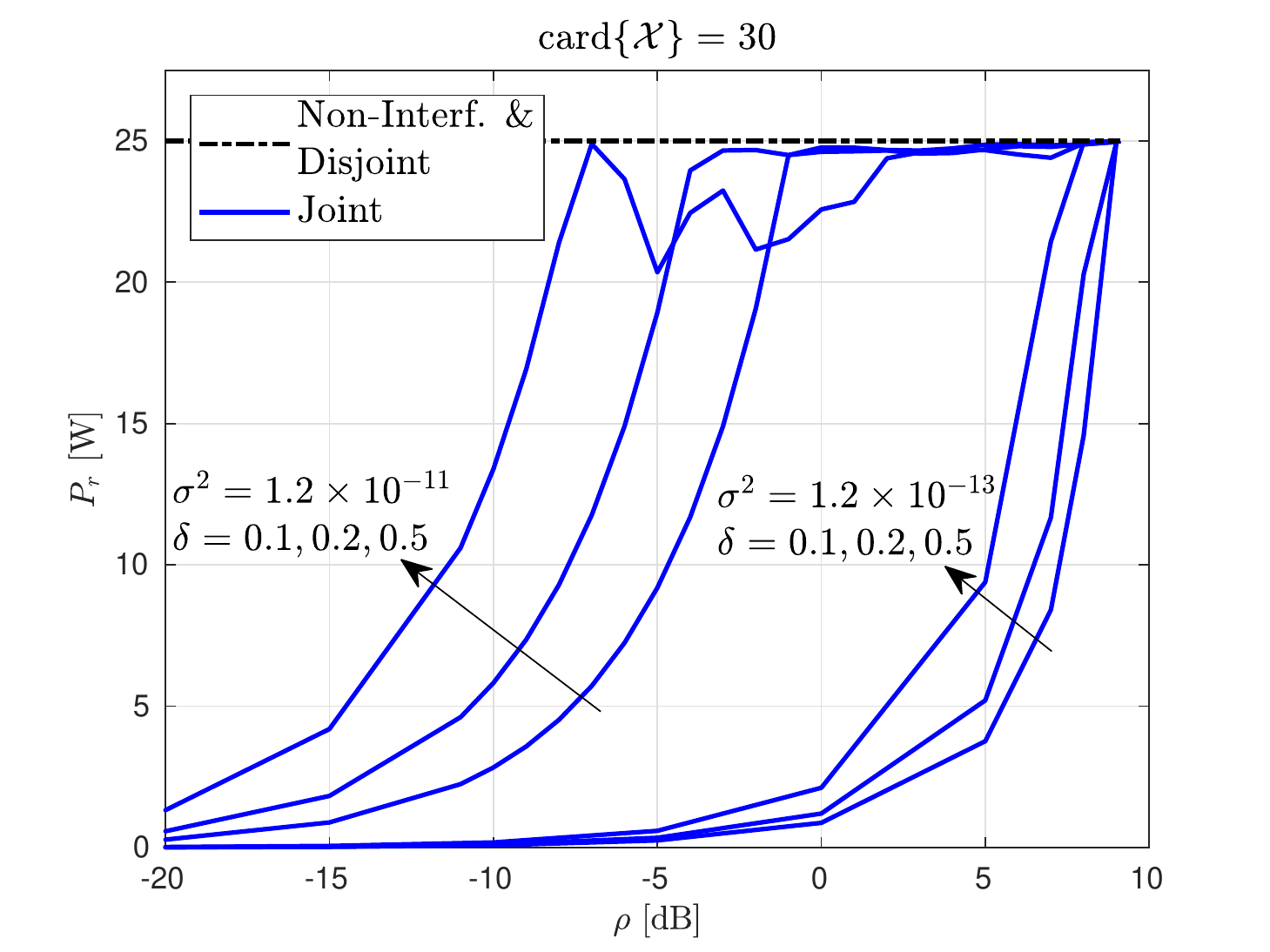}}
 \else
 \centerline{\includegraphics[width=0.75\columnwidth]{fig_08.pdf}}
 \fi
\caption{Radar transmit power versus the minimum required SDR for a joint and disjoint design, different values of the density of the interference scatterers, $\delta$, and intensity of the interference, $\sigma^2$, when 30 resolution cells are protected; for comparison purposes, the case of non-interfering systems is also included.} \label{fig_08}
\end{figure}

In Figs.~\ref{fig_07} and~\ref{fig_08}, we report, as a function $\rho$, the level of transmit communication and radar power, respectively, corresponding to the scenarios inspected in Figs.~\ref{fig_05} and~\ref{fig_06}. Clearly, when the minimum performance level required at the radar is increased, the interference becomes stronger, or the density of the interference scatterers becomes higher (large values of $\rho$, $\sigma^2$, or $\delta$, respectively), the transmit power must be increased at the radar side and decreased at the communication system side. It is interesting to notice that there are intervals of $\rho$ where both systems are transmitting at their maximum power, this meaning that coexistence is only handled by space-time beamforming at the communication transmitter and radar receiver; more generally, power control is also required to mitigate the mutual interference.

\begin{figure}[t]
 \centering
 \ifCLASSOPTIONtwocolumn
 \centerline{\includegraphics[width=1.05\columnwidth]{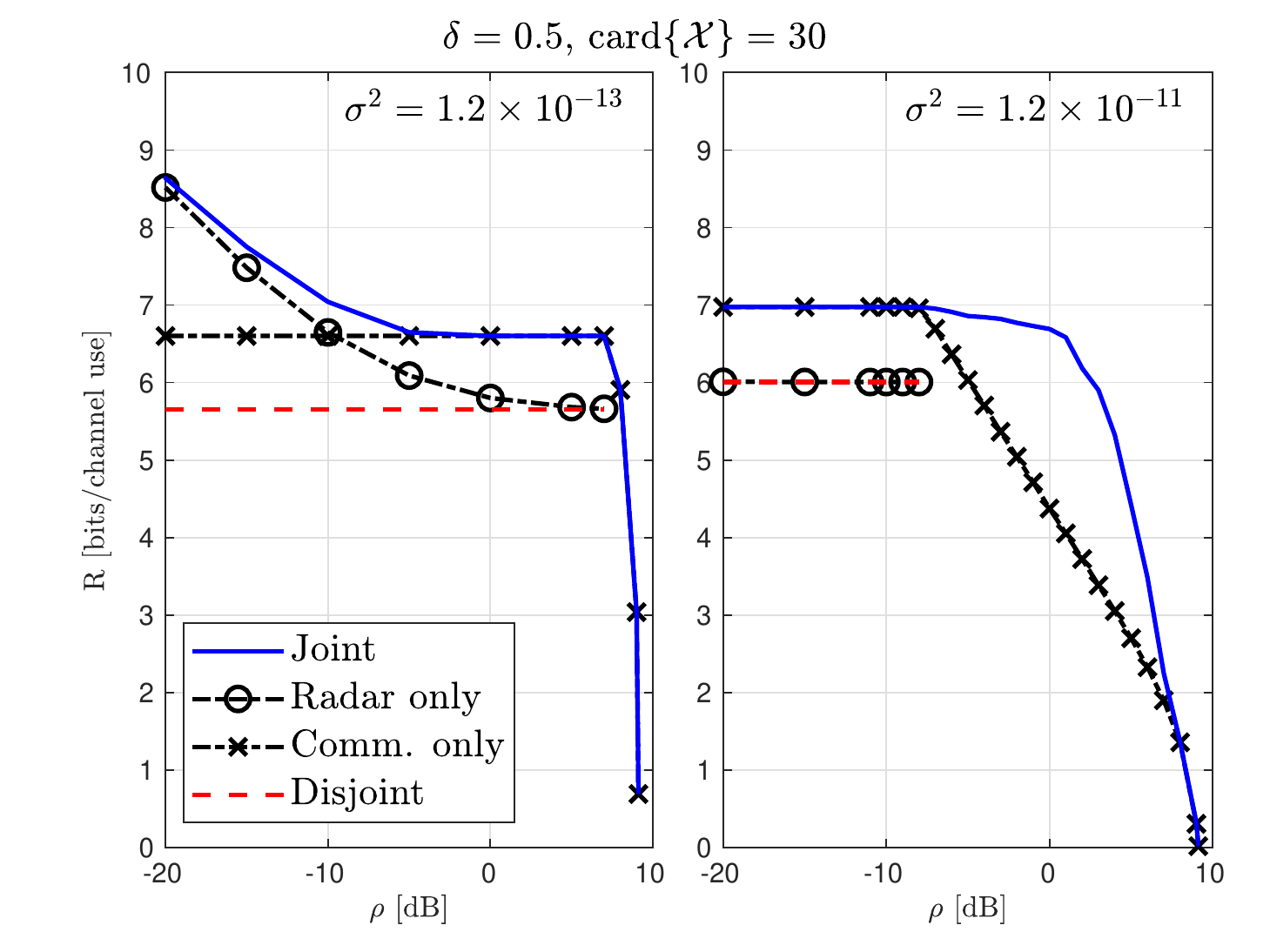}}
 \else
 \centerline{\includegraphics[width=0.75\columnwidth]{fig_09.pdf}}
 \fi
\caption{Mutual information versus the minimum required SDR for different design strategies and different values of the intensity of the interference, when the density density of the interference scatterers is $\delta=0.5$, and 30 resolution cells are protected.} \label{fig_09}
\end{figure}

Next we compare the proposed system design with two additional design strategies. In the first one, the communication system preexists the radar and ignores its presence, so that the covariance matrix of the STC's is $\bm C_w$, the waterfilling solution for the channel $\bm r = (\bm H \otimes \bm I_N) \bm c+ \bm v$. The radar is overlaid and adjusts its receive filters and transmit power so as to meet the SDR constraint and solve Problem~\eqref{opt_prob} when $\bm C=\bm C_w$. The second strategy analyzes the specular case, where the radar preexists the communication system and adopts the transmit power and receive filters in~\eqref{radar_only}, while the communication system is overlaid and solves Problem~\eqref{opt_prob} with $P_r$ and $\{w_{n,j}\}_{(n,j)\in\mathcal X}$ fixed to the values in~\eqref{radar_only}. In Fig.~\ref{fig_09}, the mutual information is reported as a function of $\rho$ for two values of $\sigma^2$ when $\delta=0.5$ and $\card\{\mathcal X\}=30$. It can be seen that, when the interference is weak, the single system optimization is almost as good as the joint design: indeed, the radar optimization exhibits a good performance when $\rho$ is small (and converges to the disjoint design when $\rho$ gets smaller), while the communication system optimiziation has nearly the same performance as the joint design when $\rho$ is large, so that the worst case loss in the mutual information is only 6\% at $-9.6$~dB. When the interference is strong, instead, the single system optimization exhibits a significant loss in a wide range of SDR of interest for the radar.

\begin{figure}[t]
 \centering
 \ifCLASSOPTIONtwocolumn
 \centerline{\includegraphics[width=1.05\columnwidth]{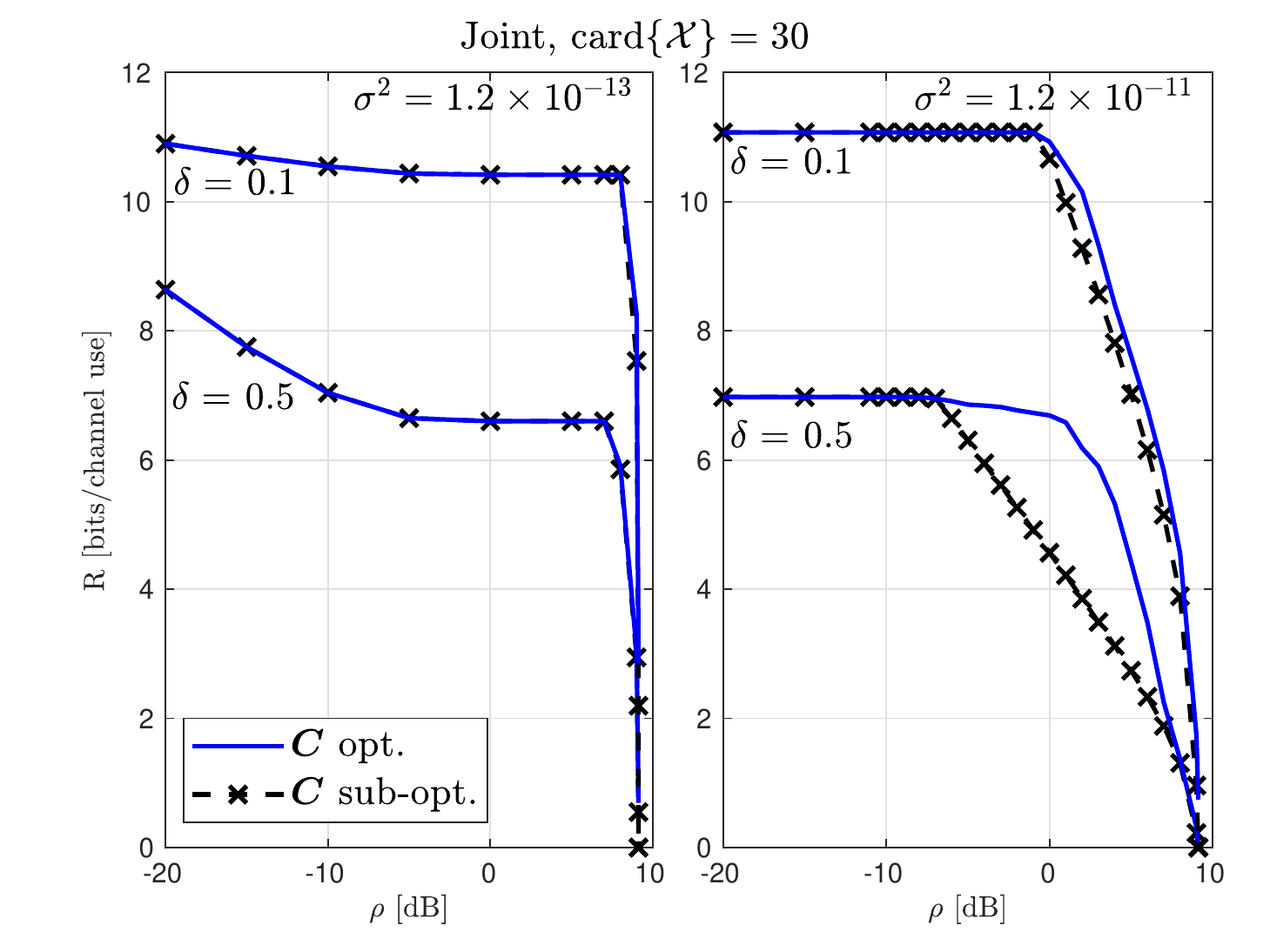}}
 \else
 \centerline{\includegraphics[width=0.75\columnwidth]{fig_10.pdf}}
 \fi
\caption{Mutual information versus the minimum required SDR for the joint design with codebook maximization problem optimally and sub-optimally solved (cfr. Secs.~\ref{opt_C_sol} and~\ref{sub-opt_C_sol}), and for different values of the intensity of the interference, $\sigma^2$, and density of the interference scatterers, $\delta$, when 30 resolution cells are protected.} \label{fig_10}
\end{figure}

In Fig.~\ref{fig_10}, we compare the solution to the codebook optimization subproblem provided in Sec.~\ref{opt_C_sol} (labeled ``$\bm C$ opt.'' in the figure) with the sub-optimum solution presented in Sec.~\ref{sub-opt_C_sol} (labeled ``$\bm C$ sub-opt.''). The mutual information is reported versus $\rho$ for two values of $\delta$ and $\sigma^2$, when $\card\{\mathcal X\}=30$. It is seen that the two solutions are almost coincident when the interference is weak. When, instead, the interference is strong, the sub-optimum solution exhibits some performance degradation for $\delta=0.1$ in the region $\rho\geq 0$~dB, and the loss becomes significant for $\delta=0.5$ and $\rho\geq -5$~dB.

\begin{figure}[t]
 \centering
 \ifCLASSOPTIONtwocolumn
 \centerline{\includegraphics[width=1.1\columnwidth]{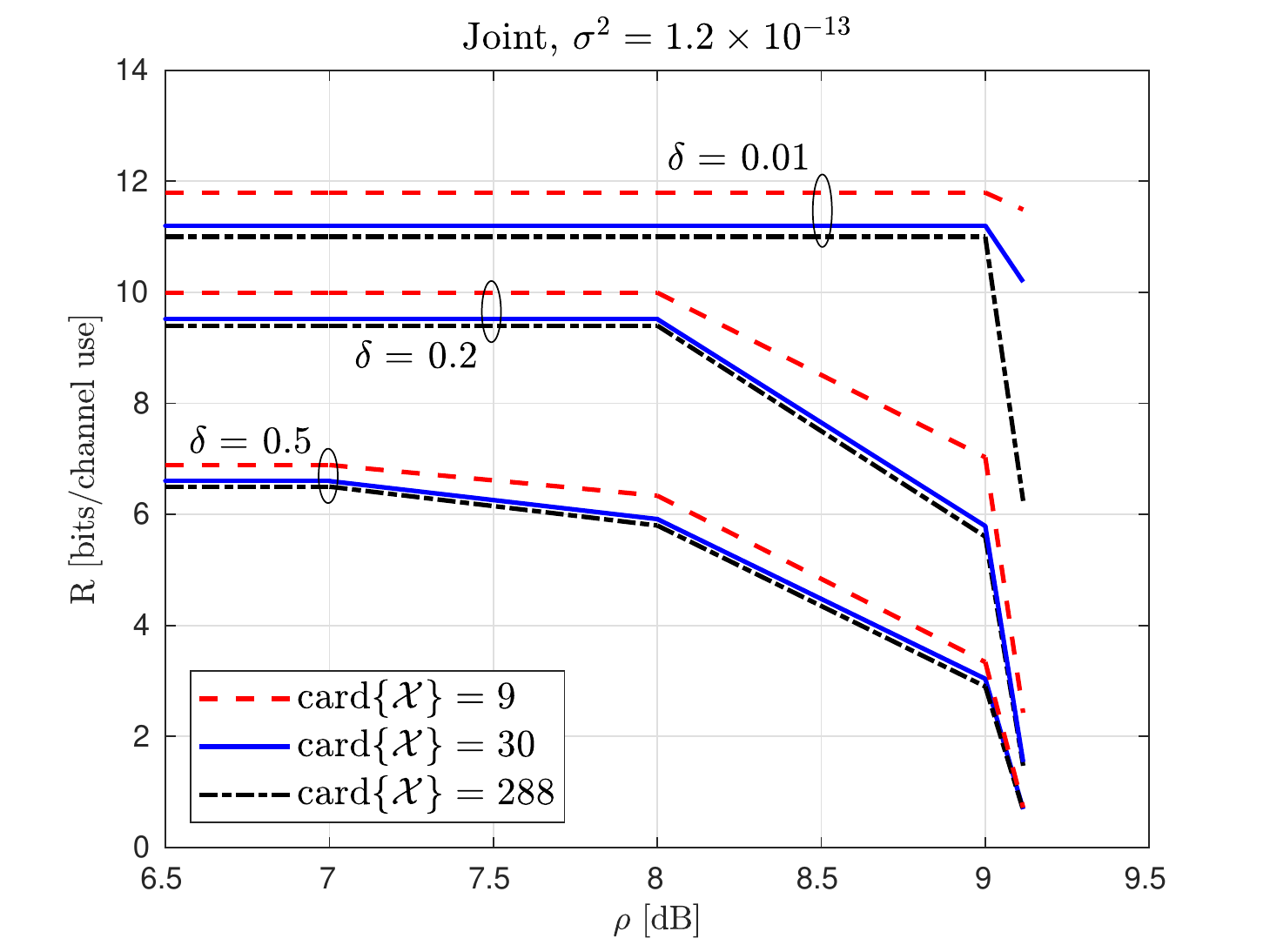}}
 \else
 \centerline{\includegraphics[width=0.75\columnwidth]{fig_11.pdf}}
 \fi
\caption{Mutual information versus the minimum required SDR in the joint design for different values of the number of protected resolution cells and of the density of the interference scatterers, $\delta$, when the intensity of the interference is $\sigma^2=1.2 \times 10^{-13}$.} \label{fig_11}
\end{figure}

\begin{figure}[t]
 \centering
 \ifCLASSOPTIONtwocolumn
 \centerline{\includegraphics[width=1.1\columnwidth]{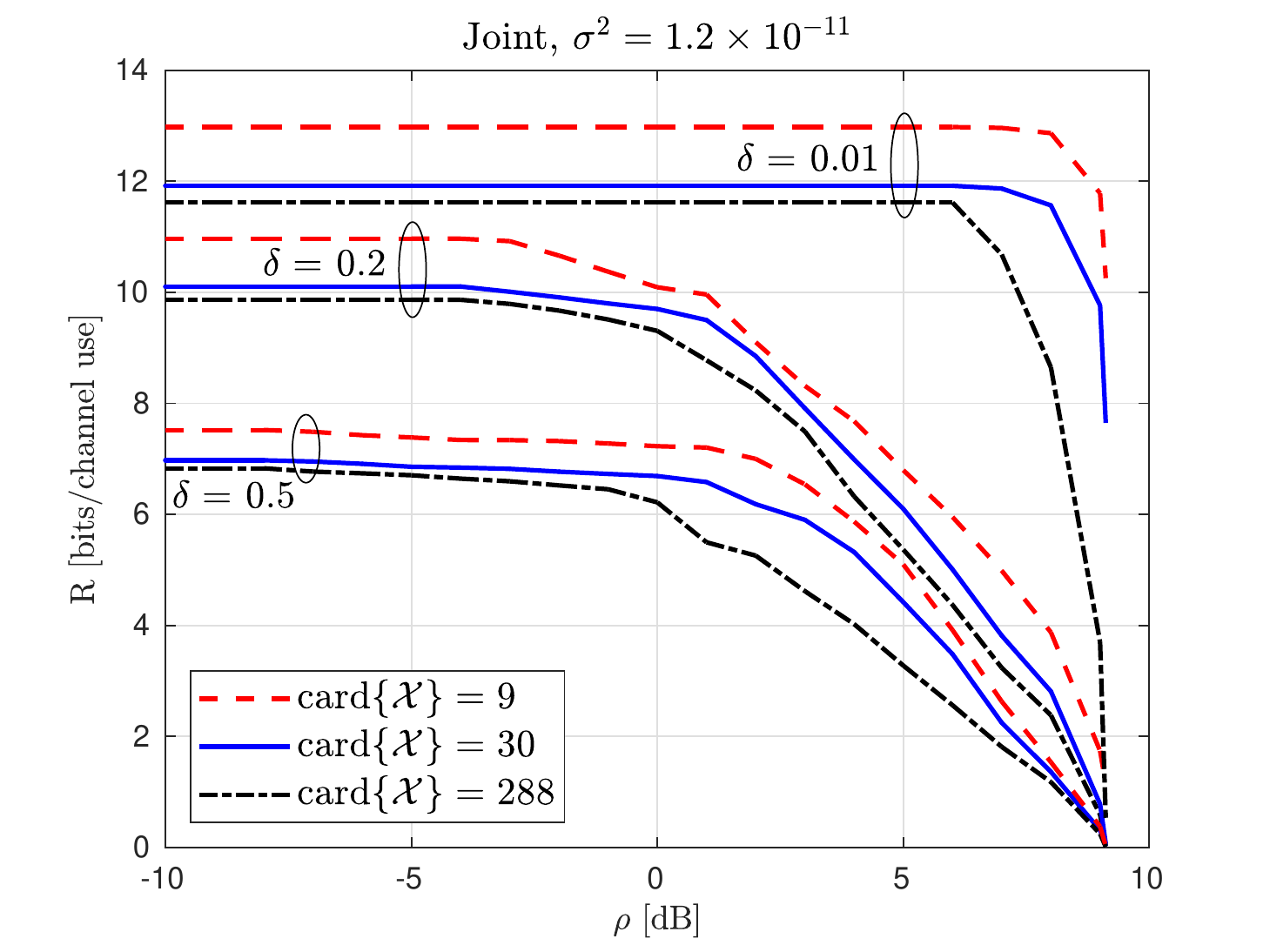}}
 \else
 \centerline{\includegraphics[width=0.75\columnwidth]{fig_12.pdf}}
 \fi
\caption{Mutual information versus the minimum required SDR in the joint design for different values of the number of protected resolution cells and of the density of the interference scatterers, $\delta$, when the intensity of the interference is $\sigma^2=1.2 \times 10^{-11}$.} \label{fig_12}
\end{figure}

Finally, we analyze the impact of the number of protected radar cells on the system performance. In Figs.~\ref{fig_11} and~\ref{fig_12}, the mutual information is reported versus $\rho$ for different values of $\card\{\mathcal X\}$ and $\delta$, when $\sigma^2=1.2\times10^{-13}$ and $\sigma^2=1.2\times10^{-11}$, respectively. Clearly, the mutual information is decreasing with the cardinality of the set of protected cells, even if, in the inspected scenario, the performance degradation rapidly saturates, and the difference between 30 and 288 (corresponding to 10.4\% and 100\%, respectively, of the total number of radar resolution cells) is significant only for small $\delta$'s and in the large $\rho$'s region.

\section{Conclusion}\label{conclusion}
In this work we tackled the problem of joint design of a radar and a MIMO communication system sharing the same bandwidth, and we proposed to maximize the mutual information with a constraint on the minimum SDR level required at each resolution cell monitored by the radar. This translates to a non-convex optimization problem, with a large number of constraints, and is sub-optimally solved by resorting to block coordinate ascent. The numerical results have shown that large gains are possible with respect to the disjoint design. Future developments will focus on the inclusion of the radar (fast-time) code in the design problem and on the generalization to STC's of the communication system that spans an integer multiple of the PRT.

\bibliographystyle{IEEEtran}
\bibliography{IEEEabrv,references}

\end{document}